\documentclass[final,3p,times,twocolumn]{elsarticle}
\usepackage[utf8]{inputenc}
\usepackage[colorlinks=true,linkcolor=black, citecolor=blue, urlcolor=blue]{hyperref}
\usepackage[dvipsnames]{xcolor}
\usepackage{xspace,soul}
\usepackage{makecell}
\usepackage{rotating}
\usepackage{ulem}
\usepackage{float}
\usepackage{amssymb}
\usepackage{amsmath}
\newcommand{\comment}[1]{\textcolor{black} {#1}}

\newcommand{\gobble}[1]{}
\makeatletter
\newcommand\newref[1]{#1\def\@currentlabel{#1}}
\makeatother 
\usepackage{lineno}

\begin{document}
\begin{frontmatter}


%
%
\title{{\bf The design and performance of the {\textit {XL-Calibur}} anticoincidence shield}}
%
%
\author[KTH1,KTH2]{N.K.\,Iyer}
\author[KTH1,KTH2]{M.\,Kiss}
\author[KTH1,KTH2]{M.\,Pearce\corref{cor1}}
\author[KTH1,KTH2]{T.-A.\,Stana}
%
%
\author[Ehime]{H.\,Awaki}
\author[WUSTL]{R.G.\,Bose}
\author[UNH]{A.\,Dasgupta} 
\author[StonyBrook]{G.\,De Geronimo}
\author[WUSTL,WUSTLM]{E.\,Gau}
\author[Osaka]{T.\,Hakamata}
\author[ISAS]{M.\,Ishida}
\author[Osaka]{K.\,Ishiwata} 
\author[Osaka]{W.\,Kamogawa}
\author[UNH]{F.\,Kislat} 
\author[RIKEN1]{T.\,Kitaguchi} 
\author[WUSTL,WUSTLM,WUSTLQ]{H.\,Krawczynski}
\author[WUSTL]{L.\,Lisalda} 
\author[ISAS]{Y.\,Maeda}
\author[Osaka]{H.\,Matsumoto}
\author[TokyoMet]{A.\,Miyamoto}
\author[OIST]{T.\,Miyazawa}
\author[Hiroshima]{T.\,Mizuno}
\author[WUSTL]{B.F.\,Rauch}
\author[WUSTL]{N.\,Rodriguez Cavero}
\author[Hiroshima]{N.\,Sakamoto}
\author[Osaka]{J.\,Sato}
\author[UNH]{S.\,Spooner}
\author[Hiroshima]{H.\,Takahashi} 
\author[TokyoMet]{M.\,Takeo}
\author[RIKEN2]{T.\,Tamagawa}
\author[TokyoSci]{Y.\,Uchida}
\author[WUSTL]{A.T.\,West}
\author[UNH]{K.\,Wimalasena}
\author[Osaka]{M.\,Yoshimoto} 

%
%

%
\cortext[cor1]{Corresponding author: M.\,Pearce ({\it pearce@kth.se}).}
\address[KTH1]{KTH Royal Institute of Technology, Department of Physics, 106 91 Stockholm, Sweden.}
\address[KTH2]{The Oskar Klein Centre for Cosmoparticle Physics, AlbaNova University Centre, 106 91 Stockholm, Sweden.}
\address[Ehime]{Graduate School of Science and Engineering, Ehime University, Bunkyo-cho, Matsuyama, Ehime, Japan.}
\address[WUSTL]{Physics Department, Washington University in St. Louis, 1 Brookings Drive, CB 1105, St. Louis, MO 63130, USA.}
\address[UNH]{Department of Physics and Astronomy and Space Science Center, University of New Hampshire, Durham, NH 03824, USA.}
\address[StonyBrook]{Stony Brook University, Stony Brook, NY 11794-2350, USA.}
\address[WUSTLM]{McDonnell Center for the Space Sciences, Washington University in St. Louis, 1 Brookings Drive, CB 1105, St. Louis, MO 63130, USA.}
\address[Osaka]{Osaka University, Department of Earth and Space Science, Graduate School of Science,
and Project Research Center for Fundamental Sciences, 1-1 Machikaneyama-cho, Toyonaka, Osaka 560-0043, Japan.}
\address[ISAS]{The Institute of Space and Astronautical Science/JAXA, 3-1-1 Yoshinodai, Chuo-ward, Sagamihara, Kanagawa 252-5210, Japan.}
\address[RIKEN1]{RIKEN Cluster for Pioneering Research, 2-1 Hirosawa, Wako, Saitama 351-0198, Japan.}
\address[TokyoMet]{Department of Physics, Tokyo Metropolitan University, 1-1 Minami-Osawa, Hachioji, Tokyo 192-0397, Japan.}
\address[WUSTLQ]{Center for Quantum Leaps, Washington University in St. Louis, 1 Brookings Drive, CB 1105, St. Louis, MO 63130, USA.}
\address[OIST]{Okinawa Institute of Science and Technology Graduate University, Kunigami-gun, Japan.}
\address[Hiroshima]{Hiroshima University, 1-3-1 Kagamiyama, Higashi-Hiroshima, Hiroshima 739-8526, Japan.}
\address[RIKEN2]{RIKEN Nishina Center, 2-1 Hirosawa, Wako, Saitama 351-0198, Japan.}
\address[TokyoSci]{Department of Physics, Faculty of Science and Technology, Tokyo University of Science, Noda city, Chiba 278-8510, Japan.}

%
%
\begin{abstract}
The {\it XL-Calibur} balloon-borne hard X-ray polarimetry mission comprises a Compton-scattering polarimeter placed at the focal point of an X-ray mirror. The polarimeter is housed within a BGO anticoincidence shield, which is needed to mitigate the considerable background radiation present at the observation altitude of $\sim$40~km. This paper details the design, construction and testing of the anticoincidence shield, as well as the performance measured during the week-long maiden flight from Esrange Space Centre to the Canadian Northwest Territories in July 2022. The in-flight performance of the shield followed design expectations, with a veto threshold $<$100~keV and a measured background rate of $\sim$0.5~Hz (20--40 keV). This is compatible with the scientific goals of the mission, where \%-level minimum detectable polarisation is sought for a Hz-level source rate.  

\end{abstract}
%
%
\begin{keyword}
Anticoincidence \sep BGO scintillator \sep Photomultiplier tube \sep X-ray polarimetry \sep scientific ballooning \sep Monte Carlo \sep qualification testing 

\end{keyword}
\end{frontmatter}
%
%
\section{Introduction}
%
\subsection{Overview}
New information on the innermost regions of black-hole and neutron-star binary systems, as well as the emission locale of isolated pulsars, can be obtained by measuring the linear polarisation of the X-ray emission, characterised through the polarisation fraction (PF, \%) and polarisation angle (PA, degrees)~\cite{Chattopadhyay.2021,Krawczynski.2011lxm,Lei.1997}. 
X-ray polarimetry is complementary to the well-established imaging, timing and spectroscopy measurements which have yielded essentially all information on such sources to date. 

\begin{figure}[tb]
\begin{center}
    \includegraphics[width=0.9\linewidth]{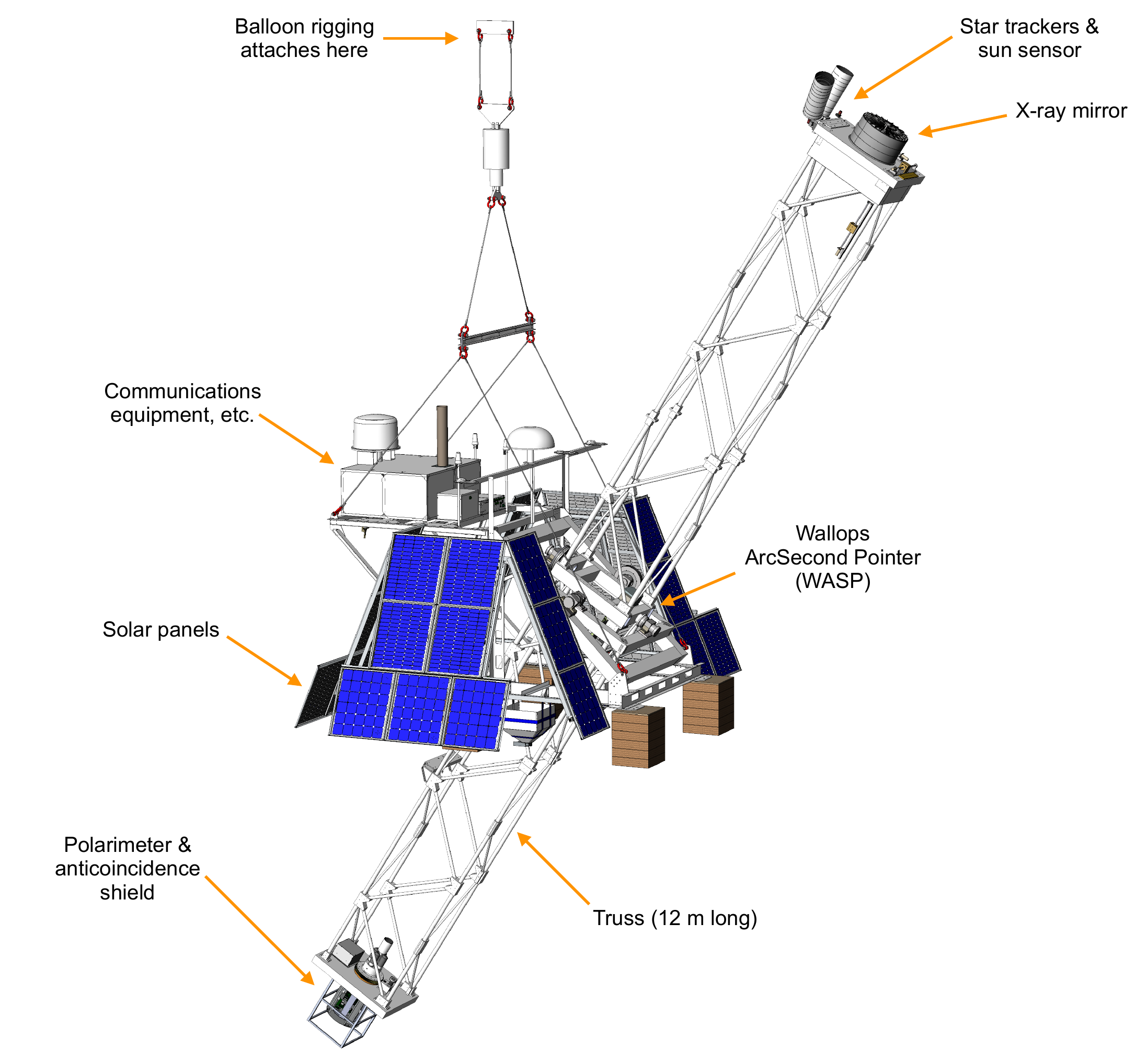}
\end{center}
\caption{\label{fig:XL-Calibur} A schematic overview of the XL-Calibur gondola. A 12~m long light-weight carbon-fibre/aluminium truss houses an X-ray mirror at one end and a polarimeter/anticoincidence assembly at the other. The truss is mounted in the Wallops Arc Second Pointer, which allows the optical axis of the X-ray mirror/polarimeter system to be pointed at celestial sources with arcsecond precision. The gondola hangs $\sim$100~m under a 1.1$\times10^6$~m$^3$ helium-filled polyethylene zero-pressure balloon. During flight, the polarimeter/anticoincidence assembly is housed inside a sheet metal box, which is covered in reflective mylar tape.}
\end{figure}

The {\it XL-Calibur}~\cite{Abarr.2021} X-ray polarimetry mission operates in the 15--80~keV energy band, which allows Compton reflection from black-hole accretion disks and cyclotron-resonant scattering in magnetised neutron stars to be studied. 
Observations are conducted from a stabilised balloon-borne platform, at an altitude of $\sim$40~km, where the atmospheric absorption is low for $>$15~keV photons. A 10~arcmin field-of-view mirror focusses X-rays over 12~m onto the polarimeter assembly, as shown in Figure~\ref{fig:XL-Calibur}. The design of {\it XL-Calibur} follows that of {\it X-Calibur}, which flew from Antarctica in December 2018~\cite{Abarr.2020p1n,Abarr.2022}. 
{\it XL-Calibur} features several improvements over {\it X-Calibur}, including a larger focal length mirror (12~m instead of 8~m), with a larger effective area; an improved anticoincidence shield; thinner and lower-background solid state Cadmium Zinc Telluride (CZT) detectors; and, lower energy threshold CZT read-out electronics.

The mirror is mounted at one end of a stiff and lightweight truss, while the polarimeter is mounted at the opposite end. The truss is pointed with arcsecond accuracy by the Wallops Arc Second Pointer (WASP)~\cite{Stuchlik.2017}. The polarimeter comprises an 8~cm long, 1.2~cm diameter, beryllium (Be) rod, which is surrounded by 4 sets of orthogonally circumadjacent CZT detectors, allowing the Compton-scattering angle of photons in an incident X-ray beam to be determined. 
Photons will preferentially scatter perpendicular to the electric field direction. 
The amplitude and phase of the resulting azimuthal counting rate modulation encodes the polarisation properties of the beam. 
\comment{The polarimeter continuously rotates about the viewing axis (twice per minute), to mitigate systematic effects arising from any non-uniform instrument response.}

Particle and photon radiation present in the stratosphere can generate a signal in the polarimeter, which cannot be distinguished from that caused by X-ray emission from a celestial source, i.e. a single energy deposit ('hit'), above threshold, in one of the CZT detectors. 
The polarimeter is therefore mounted inside the bore of a several-cm thick Bi$_4$Ge$_3$O$_{12}$ (BGO) scintillator anticoincidence shield -- the focus of this paper.  
Energy deposits in the shield volume generate a scintillation signal (referred to here as a 'veto' if above a threshold value) which allows background events to be rejected within a chosen time-window. 

{\it XL-Calibur} was launched from the Esrange Space Centre, near Kiruna, in northern Sweden on 11$^{\mathrm{th}}$ July at 23:45 UTC. The planned observations of the Crab, Cyg X-1 and Her X-1 were disrupted by technical problems during the flight. The performance of the anticoincidence shield was studied in detail, however.   

\subsection{The X-Calibur shield}
\begin{figure}[tb]
\centering
\includegraphics[width=0.5\textwidth]{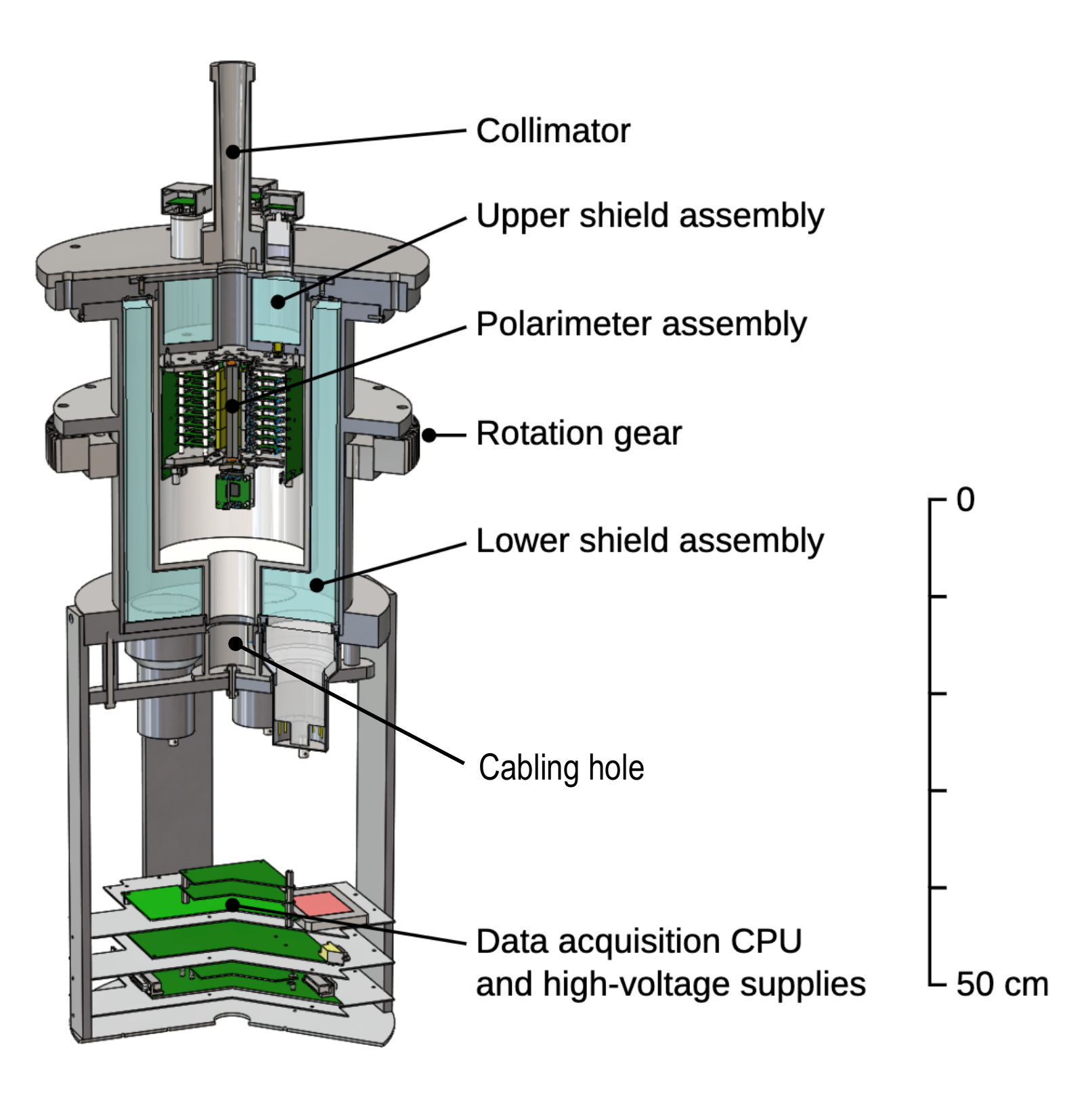}
\caption{\label{fig:XC_shield} A cross-sectional view of the the \comment{previous} {\it X-Calibur} polarimeter and anticoincidence shield.}
\end{figure}

The {\it X-Calibur} anticoincidence shield, \comment{which flew in 2018}, was constructed from CsI(Na) scintillator read out by photomultiplier tubes (PMT), as detailed in Figure~\ref{fig:XC_shield}. The volume of the shield bore needed to accommodate the polarimeter was relatively large since the CZT read-out electronics boards were arranged perpendicular to the CZT detectors. This dictated the overall dimensions of the shield (outer diameter 245~mm). The overall mass was $\sim$85~kg including the mechanics, and a 10~kg tungsten plate onto which a tungsten collimator was mounted. 

\comment{During the 2018 {\it X-Calibur} flight, a shield energy threshold of $\sim$1~MeV was needed to obtain an acceptable  polarimeter dead-time fraction $<$5\%~\cite{Abarr.2021,Abarr.2022}. 
This was significantly higher than the foreseen threshold of 100~keV. 
Laboratory tests subsequently demonstrated that the high dead-time was due to minimum-ionising particles (MIPs) passing through the 3~cm thick CsI(Na) shield. 
The resulting large-amplitude PMT pulses saturated the read-out electronics generating significant shield dead-time 
($\sim$50~$\mu$s), and consequently degraded the background rejection performance. Suppressing the amplitude of MIP-induced PMT pulses is a key requirement for the {\it XL-Calibur} shield.} 

\subsection{The XL-Calibur shield}

A reduction in the shield mass was a key goal for {\it XL-Calibur}, to offset the higher weight of the longer truss (12~m instead of 8~m), so as not to deteriorate the balloon float altitude. A shield redesign could also take advantage of the new polarimeter design for {\it XL-Calibur}, where the CZT read-out system allows a more compact design.  

The main design changes for the {\it XL-Calibur} shield are:
\begin{itemize}
\item BGO scintillator is used since (a) it has a higher stopping power (7.1~g/cm$^3$ density, compared to 4.5~g/cm$^3$ for CsI(Na)); (b) it has a faster scintillation decay time (0.3~$\mu$s, compared to 0.5~$\mu$s (slow component) and 4.2~$\mu$s (fast component) for CsI(Na)), which reduces the shield dead-time; (c) it is easier to handle, as it is not hygroscopic and is mechanically robust. 
\item Changes are made to the PMT voltage-divider circuitry, and the front-end electronics design, to mitigate saturation of the readout by MIP events and thereby achieve a 100~keV threshold in the presence of the high MIP-rate in the stratosphere.
\item \comment{Elimination of the large-diameter uninstrumented cabling aperture under the {\it X-Calibur} polarimeter (see Figure~\ref{fig:XC_shield}), which degrades the rejection performance for albedo background.}
\end{itemize}

\subsection{\comment{Measurement background and sensitivity}}
\label{sec:background}
\comment{At an atmospheric overburden of a few~g/cm$^2$,
primary (p, e, $\alpha$) and secondary (atmospheric) cosmic-rays (p, $\pi$, e, n, $\gamma$\footnote{The symbol $\gamma$ denotes both X-rays and $\gamma$-rays.}, $\mu$) generate a measurement background. 
The primary cosmic X-ray background~\cite{Ajello.2008} is attenuated by the atmosphere, resulting in flux-levels lower than the atmospheric $\gamma$-rays.} 

\comment{Charged cosmic-rays are minimum ionising and the resulting large energy deposits in the shield ($\sim$14~MeV per cm of traversed BGO) allow this background to be efficiently vetoed. 
While the low energy component of the $\gamma$-background is likely to be photo-absorbed by the shield (which may result in a veto signal), the nature of the Compton-scattering process means that the high-energy component will preferentially 
forward-scatter in the shield and may impinge a CZT detector. 
The energy deposited in the shield may be less than the veto threshold, thereby producing an irreducible background. 
Moreover, photons which lie above the polarimetry energy range when interacting in a CZT detector may be reconstructed with significantly lower energy due to the loss of charge during diffusion in the CZT bulk because of hole trapping~\cite{Abarr.2022}.} 

\comment{Neutrons produced in the atmosphere or in detector materials also contribute significantly to the background rate. Neutrons can traverse the shield without depositing energy and impinge a CZT detector.
Simulations have shown the contribution to the polarimeter background rate is comparable to that from  
$\gamma$-rays~\cite{Abarr.2021}. Compared to hard X-ray polarimeters which utilise low atomic number detectors, e.g. plastic scintillator~\cite{Chattopadhyay.2021}, the high effective atomic number of CZT reduces the importance of elastic scattering processes. 
The production of multiple prompt $\gamma$-rays in a characteristic of processes such as de-excitation of Cd after neutron capture (with a high cross-section $<$0.4~eV~\cite{Bloser.1998t4q}); the decay of isomeric states produced when a neutron interacts with Cd, Zn or Te, or when charged particles interact with Bi or Ge; and inelastic processes such as (n,n$^\prime$,$\gamma$). These backgrounds are suppressed since the multiple $\gamma$-rays are likely to be detected by the shield.} 
\comment{An irreducible background may arise from neutron reactions which produce radioactive isotopes (activation) in material in the vicinity of the shield/polarimeter. Isotopes with long decay times compared to the anticoincidence time-window may yield delayed gamma-rays which cannot be vetoed.}

\comment{An intrinsic source of high-energy background is the decay of $^{40}$K ($\beta$-radiation (maximum energy 1.33~MeV) and $\gamma$-rays (1.46~MeV)), present in building materials on ground and the PMT glass window.}

\comment{At rigidities below $\sim$1~GV, primary charged cosmic-ray spectra (and consequently also the secondary spectra) depend on the geomagnetic latitude, $\lambda$, at which observations are conducted. The background is therefore lower at Esrange ($\lambda\sim65^\circ$) than at McMurdo ($\lambda\sim80^\circ$) on Antarctica (foreseen for future launches).
Background levels also depend on the 11-year solar activity cycle, with the lowest background fluxes present during solar-maximum conditions (predicted to next occur in $\sim$2025).}

\comment{For given background conditions, the measurement sensitivity can be expressed in terms of the Minimum Detectable Polarisation (MDP)~\cite{Weisskopf.2010855k}, where there is a 1\% chance of measuring a polarisation fraction which exceeds the MDP for an unpolarised beam. The MDP is defined as
\begin{equation}
    \mathrm{MDP} = \frac{429\%}{\mu R_s}\sqrt{\frac{R_s + R_b}{t_{\mathrm{obs}}}}, \label{eqn:MDP}
\end{equation}
where $\mu$ is the modulation factor (which describes the polarimetric response~\cite{Chattopadhyay.2021}), $R_s$ ($R_b$) is the source (background) rate in Hz, and $t_{\mathrm{obs}}$ is the duration of the source observation in seconds.
The signal rate is a few~Hz for a 1~Crab source (depends on the source elevation and balloon altitude). 
Observing the Crab for 6~hours each day during a week-long flight results in \%-level MDP if R$_\mathrm{b}$ $<$ 1~Hz. This MDP is an order-of-magnitude lower than previous missions in the hard X-ray band.
The R$_\mathrm{b}$-value for a bare polarimeter is a few hundred Hz. 
The shield is designed to reduce this background rate by two orders-of-magnitude.}

%
%
\section{Design requirements}
\label{sec:desreq}
Monte Carlo simulations of the stratospheric background have been used to study the impact of the shield design on the polarimeter background rate, $R_b$.   
Design choices were constrained by the requirement to reuse electrical and mechanical interfaces from the preceding {\it X-Calibur} mission. Table~\ref{tab:req} lists the design requirements. 
\begin{table*}[h!]
\linespread{1}\small
    \caption{Requirement checklist for the anticoincidence shield.}
    \label{tab:req}
    \begin{tabular}{|l|l|p{3.5cm}|p{5cm}|}
        \hline
        {\#} & {Parameter} & {Requirement} & {Remarks} \\
        \hline
        \multicolumn{4}{|l|}{\it \bf I. Scientific requirement} \\ 
        \hline
        \newref{I-a.}\label{req:bgrate} & Background rate & $<$1~Hz & to achieve target MDP (Equation~\ref{eqn:MDP}). \\ 
        \hline
        \multicolumn{4}{|l|}{\it \bf II. Shield detector requirements} \\ 
        \hline
        \newref{II-a.}\label{req:thresh} & Veto threshold & $\leq$100~keV & from Monte Carlo simulations (Fig.~\ref{fig:simulation-res}) \\ [0.5ex]
        \newref{II-b.}\label{req:vetorate} & Anticoincidence trigger rate limit & 20 kHz & See \S\ref{sec:simulation}.\\ [0.5ex]
        \newref{II-c.}\label{req:ltleak} & Light tightness & Required. & Light leaks impair low-energy sensitivity.  \\ [0.5ex]
        \newref{II-d.}\label{req:thermal} & Temperature range & $-$35 $^\circ$C to 30~$^\circ$C & Due to varying solar illumination during flight. \\ [0.5ex]
        \newref{II-e.}\label{req:vac} & Operating pressure & 4--1000~mbar & Prevent high-voltage discharge at low pressure. \\ 
        \newref{II-f.}\label{req:led} & Response calibration & For pre-flight testing only & Should not rely on radioactive sources, since this may not be feasible at all launch sites.\\
        \newref{II-g.}\label{req:thrscan} & Response monitoring & For use pre-flight and in-flight & Allow measurement of background rate and shield response. \\
        \newref{II-h.}\label{req:kalpha} & Fluorescence line suppression & -- & BGO fluorescence emission (28~keV, 30~keV) should not enter the polarimeter. \\
        \hline
        \multicolumn{4}{|l|}{\it \bf III. Thermo-mechanical requirements} \\
        \hline
        \newref{III-a.}\label{req:mechfit} & Interface constraints & Compatibility with existing {\it X-Calibur} mechanics, including roll bearing & See \S\ref{sec:mechanics}.\\ [0.5ex]        
        \newref{III-b.}\label{req:vibrn} & Vibration tolerance & Protect BGO crystal and PMTs from damage & Pre-flight transport and parachute landing after flight. \\ [0.5ex] 
        \newref{III-c.}\label{req:focii} & Focal-point placement & Polarimeter must be located 12~m from mirror & Be rod must lie in the focal plane.
        \\ [0.5ex]
        \newref{III-d.}\label{req:thermal2} & Thermal management of polarimeter & A thermal path should be provided from the polarimeter so that the operating temperature range is not exceeded & Polarimeter is housed in a copper can and has an electrical power rating of 12~W, with up to 49~W of heating power provided by in-built heaters.   \\ [0.5ex]      
        \newref{III-e.}\label{req:mass} & Mass & The overall mass of the shield assembly should be $<$85~kg & Mass limit corresponds {\it X-Calibur} shield solution \\
        \hline
        \multicolumn{4}{|l|}{\it \bf  IV. Electronics requirements} \\
        \hline
        \newref{IV-a.}\label{req:deadmitigate} & Mitigate effect of MIPs & Reduce the 50~$\mu$s shield dead-time seen for \it{X-Calibur} & See \S\ref{sec:simulation} and \S\ref{sec:daq}. \\ [0.5ex]
        \newref{IV-b.}\label{req:dtime} & Polarimeter dead-time measurement & $\sim$1~$\mu$s accuracy & Measurement of shield-veto induced polarimeter dead-time is required for accurate source flux analysis. \\ [0.5ex]
        \newref{IV-c.}\label{req:elefit} & Interface constraints & Compatibility with with existing {\it X-Calibur} electronics &  See \S\ref{sec:daq}.\\
        \hline
    \end{tabular}
\end{table*}
%

\section{Monte Carlo simulations}
\label{sec:simulation}
\begin{figure}[tb]
    \centering
    \includegraphics[width=0.5\textwidth]{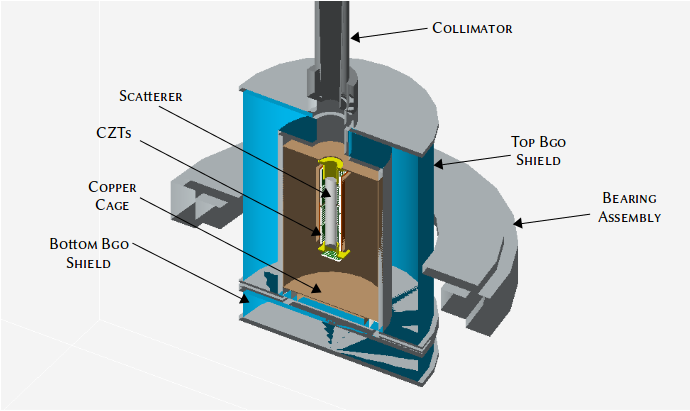}
    \caption{\label{fig:model} The geometry used for in the simulation studies. 
    The PMTs and CZT electronics are not included. The bearing assembly is mounted on a light-weight aluminium-composite honeycomb panel (shown in Figure~\ref{fig:XL-Calibur}), which is also not included.
    }
\end{figure}

A variety of shield designs were studied using the Geant4 framework~\cite{Allison.2016,Agostinelli.2003}. 
A typical simulation model is shown in Figure~\ref{fig:model}.

Particles are directed onto a geometric model of the polarimeter housed within the shield.  The shield comprises an inverted BGO well seated on a solid disc onto which the polarimeter is mounted. Polarimeter cables exit the assembly radially at the interface between the two shield halves. BGO scintillators are housed in 5~mm thick aluminium enclosures. The polarimeter is represented as a Be rod surrounded by CZT detectors, which are mounted on 3~mm thick copper heat sinks. The polarimeter is mounted in a cylindrical copper Faraday cage with wall thickness 0.8~mm. 
The model includes high-density items in the vicinity of the polarimeter, such as the tungsten collimator and the gear-wheel/bearing assembly.
The bearing attaches to the shield mechanics, allowing the polarimeter to rotate around the viewing axis during observations. 
The PMTs, CZT front-end electronics, and light-weight aluminium-composite honeycomb panel are not included.
Simulations were conducted for a shield/polarimeter elevation of 35$^\circ$ (the mean elevation of the Crab viewed from Esrange). Simulations conducted for an elevation of 55$^\circ$ gave comparable results, due to the dominantly albedo nature of the background. 

The particle background spectra and angular distributions are obtained from MAIRE~\cite{Lei.2006}\footnote{\url{https://www.radmod.co.uk/maire}} and interactions are governed by the Geant4 'shielding' physics list\footnote{\url{https://www.slac.stanford.edu/comp/physics/geant4/slac_physics_lists/shielding/shielding.html}} with the inclusion of the Livermore low energy electromagnetic physics list. The step length inside the CZT pixels was reduced to 0.05~mm to account for charge splitting between pixels. 
The simulated particle flux comprises up- and down-going atmospheric electrons, neutrons and photons, up-going atmospheric protons, and down-going atmospheric and primary protons. The spectra are detailed elsewhere~\cite{Abarr.2022}.
Simulations assume constant geomagnetic cut-off for the location of Esrange. In MAIRE, solar activity (modulation potential, $\Phi$) is specified at run-time by selecting an appropriate date. A date from the previous solar cycle was selected, \comment{14$^{\mathrm{th}}$ April, 2014}, which had the same level of solar activity (sun-spot number) as that measured at the time of the XL-Calibur flight. 
\begin{figure}[tb]
    \centering
    \includegraphics[width=0.5\textwidth]{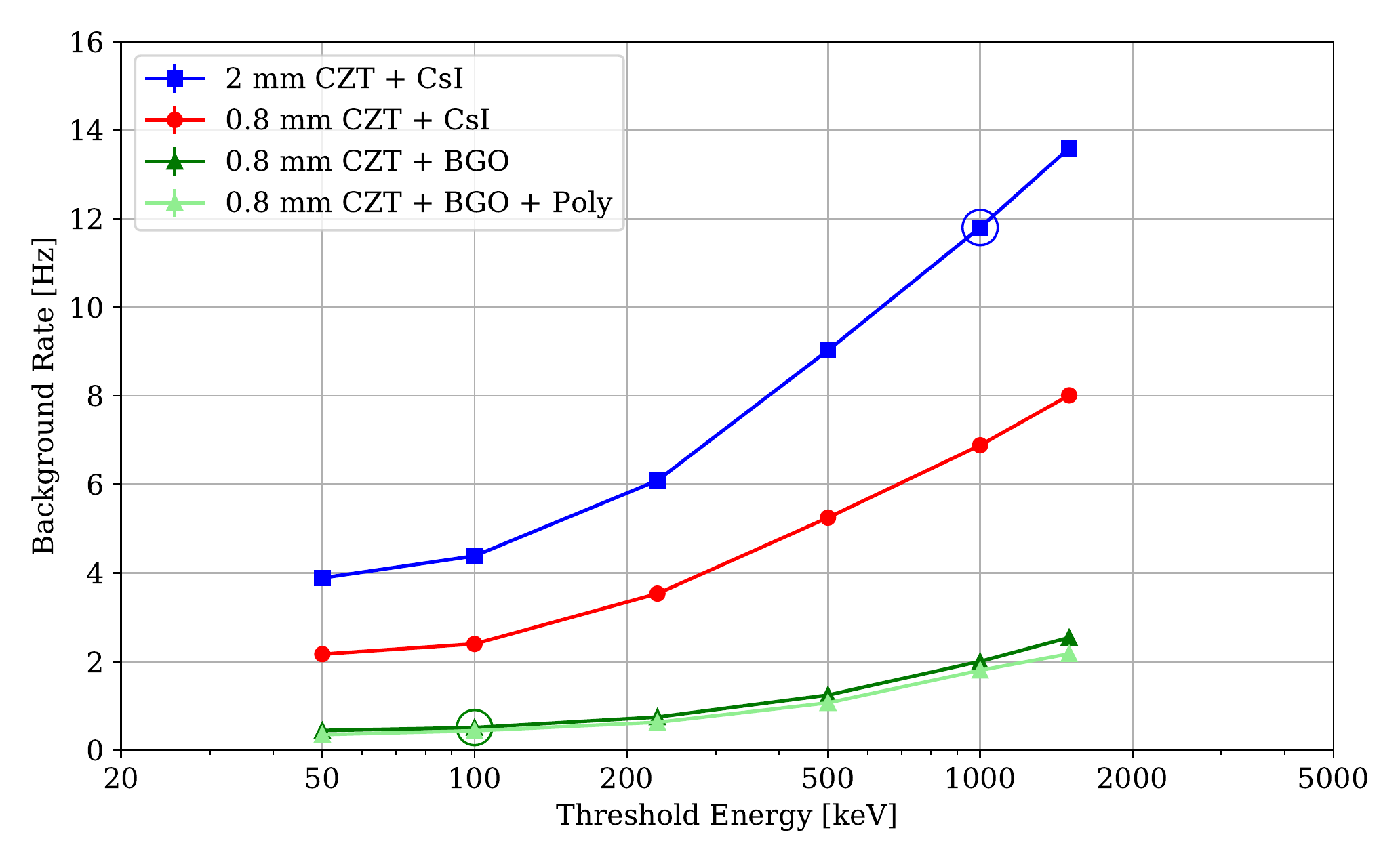}
    \caption{\label{fig:simulation-res}  Simulated polarimeter background rates (15--80~keV) for different shield configurations, with solar activity conditions assumed for July 2022. 
    The {\it X-Calibur} (dark blue squares) flight configuration \comment{(top/bottom wall thickness 30~mm, and side wall thickness 40~mm)} and that predicted for \textit{XL-Calibur} (dark green triangles) are marked with circles.
    }
\end{figure}

\begin{figure}[tb]
    \centering
    \includegraphics[width=0.5\textwidth]{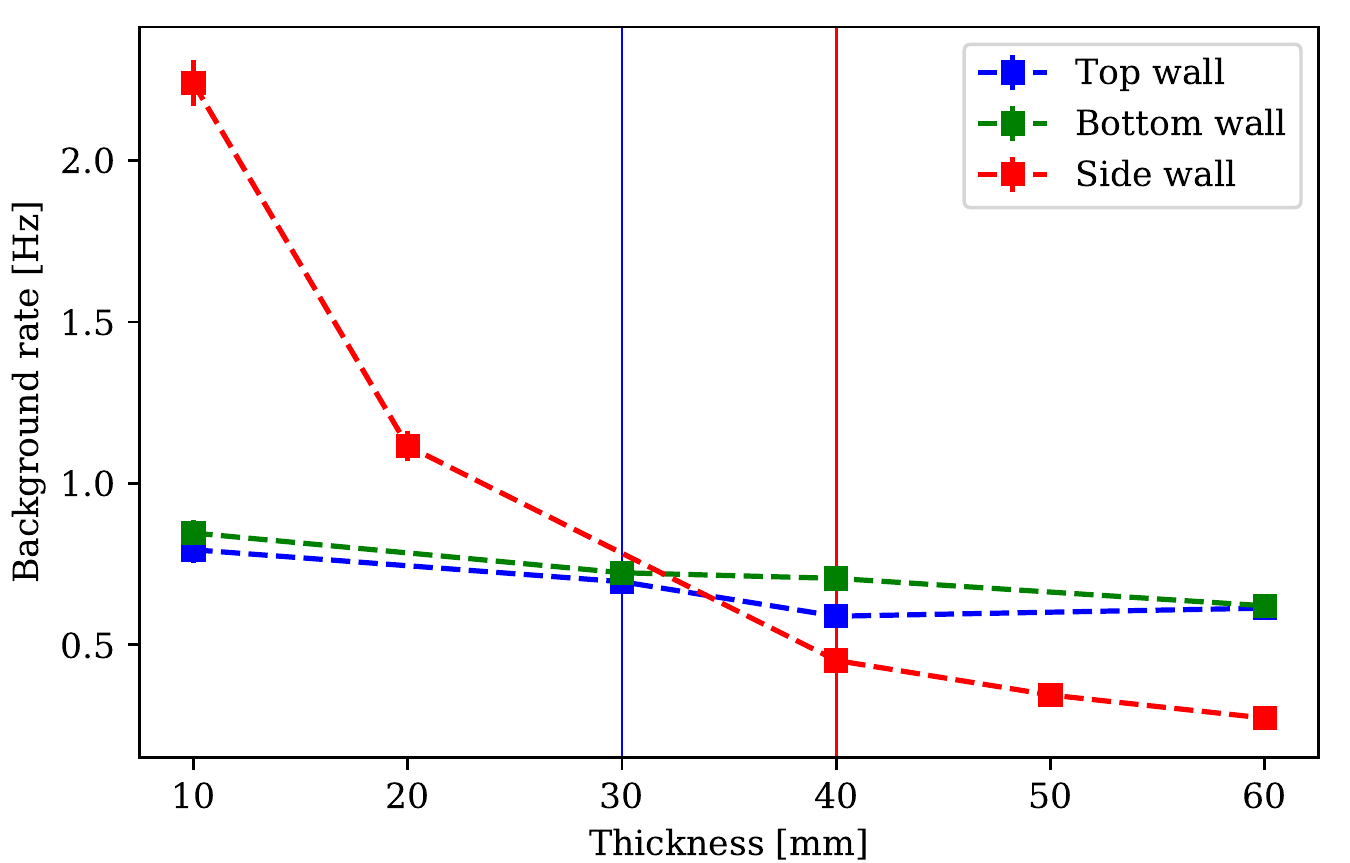}
    \caption{\label{fig:simulation-geom} Simulated polarimeter background rates with different BGO wall thickness. Thicknesses above 50 mm increase the crystal mass beyond 85~kg, exceeding the mass constraints. The finalised thickness values for top/bottom wall (at 30 mm in blue) and side wall (at 40 mm in red) are indicated by coloured vertical lines. \comment{Only one wall thickness is changed at a time, and the remaining wall thicknesses are set to 30~mm. A 100~keV threshold is applied in all cases.}}
\end{figure}

Due to redesigned read-out electronics~\cite{Abarr.2021}, the {\it XL-Calibur} polarimeter occupies a smaller volume than {\it X-Calibur}. Consequently, the inner wall of the shield can lie closer to the polarimeter, and the shield volume is smaller. Moreover, 
{\it XL-Calibur} uses thinner CZT detectors (0.8~mm instead of 2~mm), which reduces the background rate by a factor of $\sim$2. 
A variety of shield geometries (e.g. different shield wall thicknesses) were studied.
The effect of an additional high density polyethylene layer surrounding the shield (to suppress the background from atmospheric neutrons), and the background dependence on the shield veto threshold and dead-time were also studied.
Primary conclusions from these studies are summarised below and illustrated in Figure~\ref{fig:simulation-res} and Figure~\ref{fig:simulation-geom}.
\begin{itemize}

    \item Surrounding the polarimeter with $\sim$3~cm BGO wall thickness results in a background rate of a few~Hz (Fig.~\ref{fig:simulation-geom}), while still keeping the overall anticoincidence mass ($\sim$45~kg) and longest dimension 
(\mbox{$\lesssim$30~cm}) within limits placed by the \textit{X-Calibur} design. Additional passive absorbing material (such as the tungsten top-plate used in {\it X-Calibur}) offered no significant improvement in background rejection performance. 

    \item Such a BGO shield rejects $>$99\% of the charged particle (electrons and protons) background.

    \item High energy ($>$10 MeV) albedo gamma-rays and neutrons are primarily responsible for the background which leaks through the shield without raising a veto signal. 

    \item Reducing the veto energy threshold below 100~keV has limited effect, as seen in Fig.~\ref{fig:simulation-res} (requirement~\ref{req:thresh}).
 
    \item The shield veto rate reduces from $\sim$50~kHz for {\it X-Calibur} to $\sim$20~kHz for {\it XL-Calibur} -- see requirement~\ref{req:vetorate} -- thereby reducing the polarimeter dead-time. 

    \item The BGO shield passively reduces the background in the polarimeter \comment{from a few hundred Hz to $\sim$30~Hz}, and actively reduces the background by two orders of magnitude (to give a background-induced rate of $\sim$ 0.5 Hz). A fully efficient active veto is therefore a very important part of the shield design.  

    \item The addition of a 3~cm thick layer of polyethylene provides a marginal reduction in the neutron background since most of the slow neutrons stopped by the polyethylene are also stopped or flagged by the BGO in (n,$\gamma$) reactions. A thicker polyethylene layer would lead to an unacceptable increase in the payload mass and is not compatible with the current mechanical interface. 

\end{itemize}

%
%
\section{Shield design}
\label{sec:design}
%
\subsection{Overview}
\begin{figure*}[tb]
\begin{center}
    \includegraphics[width=0.9\linewidth]{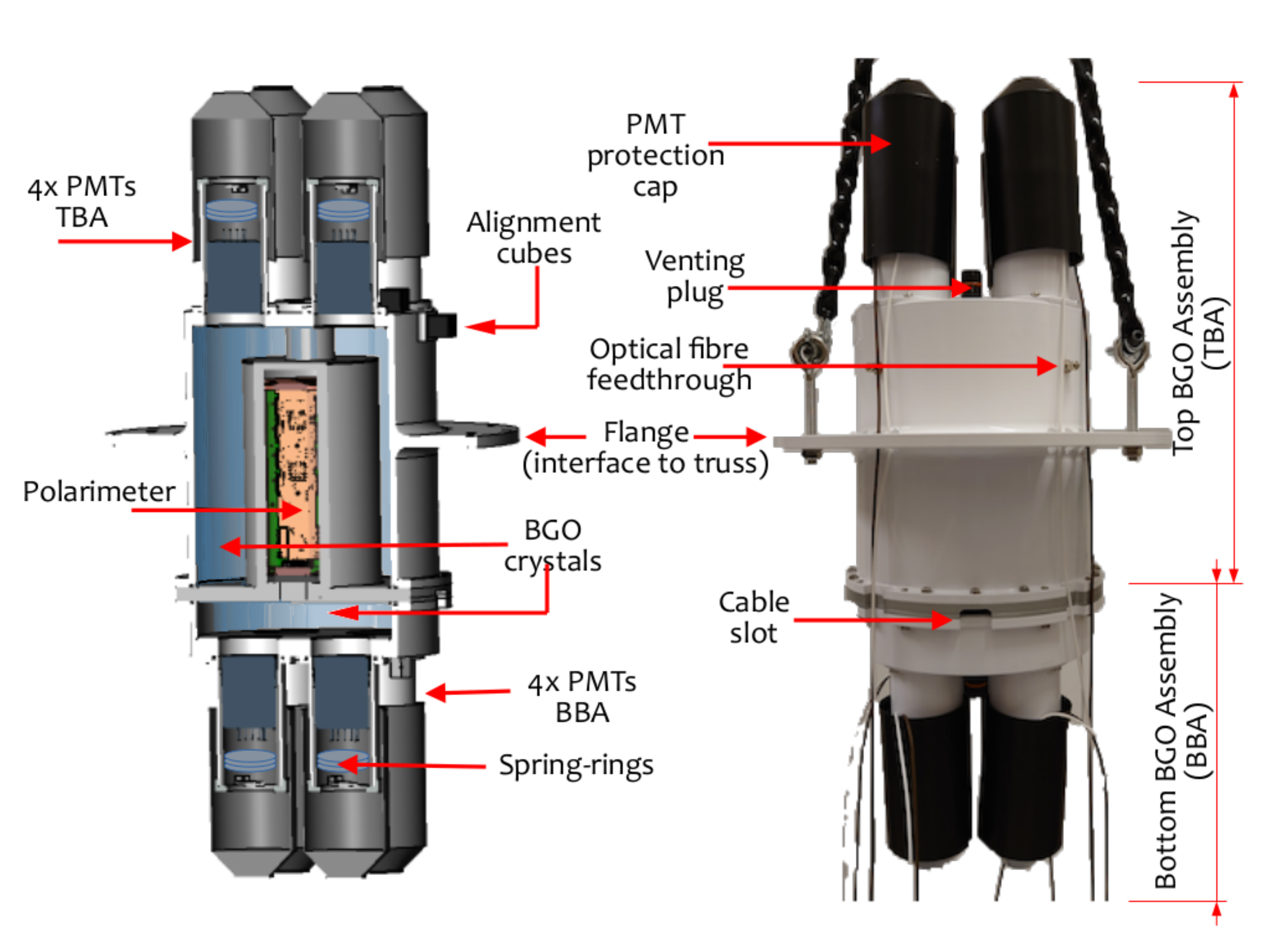}
\end{center}
\caption{\label{fig:overview} Overview of the TBA and BBA assemblies. The left-hand image shows a cross-section through the CAD model. A photograph of the shield is shown on the right-hand side. The height of the full assembly is $\sim$70~cm, with a total weight of $\sim$60~kg.}
\end{figure*}

The anticoincidence shield comprises two assemblies, as shown in Figure~\ref{fig:overview}. The polarimeter is mounted on BBA (Bottom BGO Assembly), with the aluminum casing providing a heat path from the polarimeter. The TBA (Top BGO Assembly) is formed as an inverted well, into which the polarimeter is inserted. The overall weight of TBA (BBA) is 45.5~kg (12.0~kg).     
With the exception of four small slots equispaced around the TBA/BBA interface, the resulting assembly fully surrounds the polarimeter and shields background from all directions. A 20~cm long, 8-10~mm thick tungsten collimator is mounted at the shield aperture to provide a field-of-view comparable to the mirror point-spread function ($\sim$2$'$ Half Power Diameter), thereby reducing bore-sight background. 
The components of the shield are described in the following Sections.

\subsection{Mechanical assembly}
\label{sec:mechanics}
The shield mechanics (PMT housing) is made from 5~mm (3~mm) thick aluminium alloy 6061.
The shield wall thickness prevents Ge K$_{\alpha,\beta}$ and Bi L$_{\alpha,\beta}$ BGO fluorescence lines~\cite{Furuta.2015} 
from impinging the polarimeter CZT detectors (requirement~\ref{req:kalpha}).
A flange machined into TBA interfaces to a ring bearing assembly mounted on the aluminium-composite honeycomb panel, which is mounted at the end of the truss. This allows the polarimeter and shield to rotate around the viewing axis during observations. The mechanics wall thickness is dimensioned to maintain the polarimeter alignment with the mirror and to protect components during the shock-levels expected during parachute deployment and landing at the end of the flight.
Silicone O-rings are used to prevent light leakage (requirement~\ref{req:ltleak}) through joints between mechanical elements. 
Light-tight vents are mounted on TBA and BBA to ensure that the interior of the shield follows the ambient pressure.
Optical fiber feedthroughs (one for each quadrant of TBA and one for BBA) allow LED calibration of the PMT and readout electronics (requirement~\ref{req:led}). 
In order to dampen shocks and vibrations during transport and launch/landing operations, 1-2~mm thick silicone sheets are inserted between
the BGO crystal face and the interior wall of the mechanics (requirement~\ref{req:vibrn}). 
 
\subsection{BGO scintillator}
\label{sec:bgo}
\begin{figure}[tb]
    \centering
    \includegraphics[width=.475\textwidth]{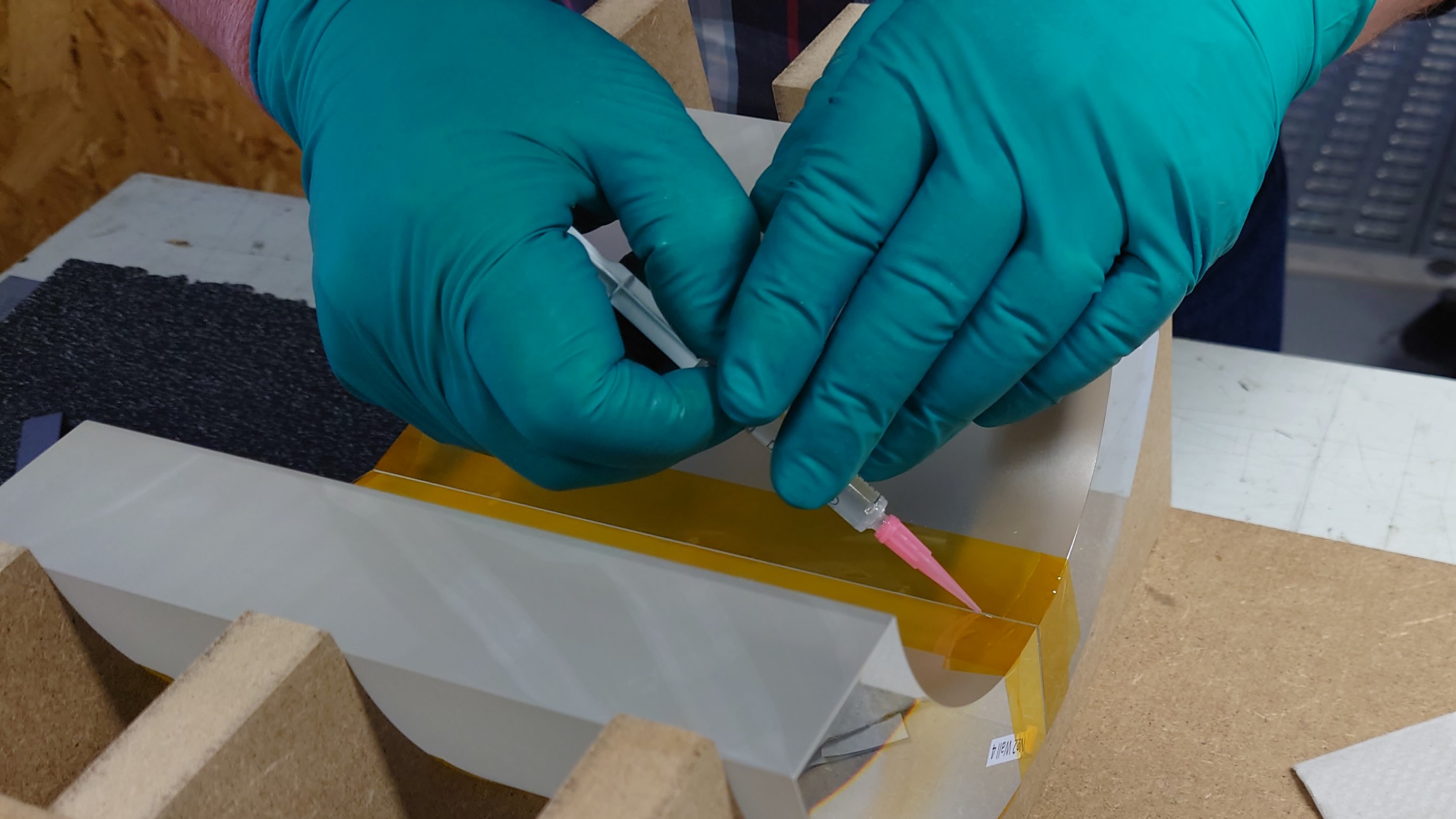}
    \includegraphics[width=.475\textwidth]{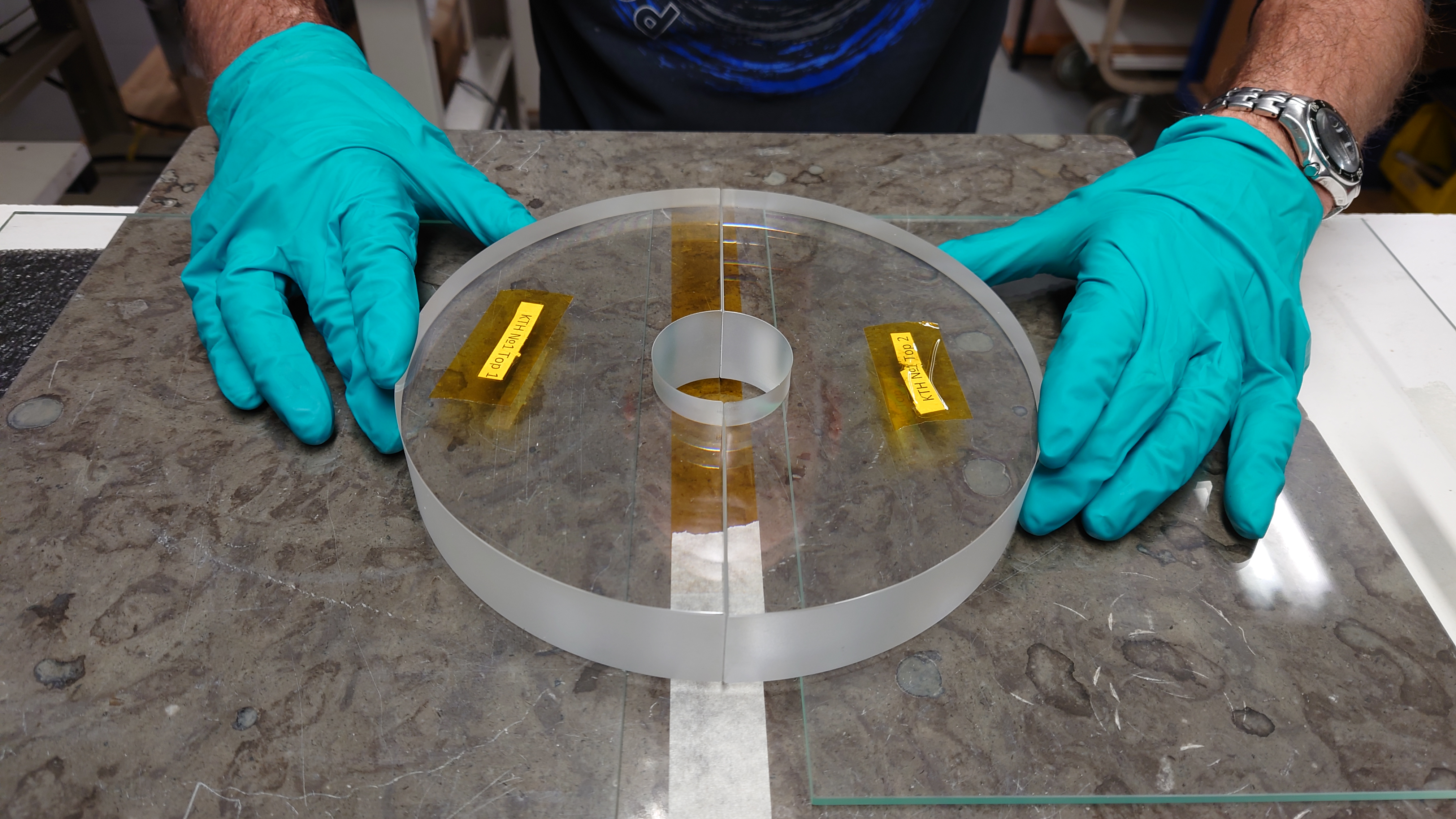}
    \caption{\label{BGO gluing pictures} \comment{\bf Left:} Two of the side TBA scintillator elements (196~mm long) being glued together. The crystals were placed into contact and the epoxy adhesive applied along the joint, allowing capillary action to draw the adhesive into the joint. \comment{\bf Right:} The two halves of the TBA top piece (overall diameter 203~mm) shown during the gluing process. X-rays passing through the collimator reach the polarimeter via the hole. The side wall thickness is \mbox{4 cm} while the lid is \mbox{3 cm} thick, for a crystal weight of \mbox{$\sim$35 kg}. 
The BBA scintillator assembly (not shown) has the same shape and segmentation as the lid, but without the hole, and it weighs \mbox{$\sim$7 kg}.}
\end{figure}

The BGO scintillators were originally procured for the PoGOLite mission~\cite{Chauvin.2016} from Nikolaev Institute of Inorganic Chemistry\footnote{\url{http://niic.nsc.ru/institute/881-niic}} in 2007. Spare material was returned to NIIC in 2020, melted down, and used to grow the {\it XL-Calibur} shield crystals.
Due to manufacturing constraints on the crystal boule size, the TBA (BBA) crystal is segmented into 4 (2) pieces, as shown in Figure~\ref{BGO gluing pictures}.

The effect of different crystal surface treatments on light-yield was evaluated by the manufacturer using scaled-down crystals.  
Flat crystal surfaces are polished to a mirror-like finish and curved surfaces are chemical etched. 
Each delivered crystal was checked for dimensional conformity, internal defects, and characterised for light-yield uniformity using a flight PMT (Section~\ref{sec:pmt}) with \mbox{$^{241}$Am} (59.5~keV) and \mbox{$^{137}$Cs} (662~keV) radioactive sources.
The BGO crystals were glued together using Epoxy Technology EPO-TEK 301\footnote{\url{https://www.epotek.com/docs/en/Datasheet/301.pdf}} at Stockholm Polymerteknik AB\footnote{\url{https://polymerteknik.com/}}.  

After gluing, the TBA and BBA crystal assemblies were wrapped in double layers of ESR sheets\footnote{Enhanced Specular Reflector, manufactured by 3M.}, which reflect $>$98\% of the scintillation light over a large range of incidence angles. 

\subsection{Photomultipliers}
\label{sec:pmt}
Both TBA and the BBA are redundantly read out using four Hamamatsu R6231-100-01/001 PMTs, as shown in Figure~\ref{fig:overview}. The devices comprise a glass vacuum tube, which has a window diameter of 51~mm (with a minimum photocathode diameter of 46~mm), and a body length of 77~mm. All PMTs are screened by the manufacturer to have a peak quantum efficiency of at least 38\%. The quantum efficiency exceeds 10\% in the range  
$\sim$300-550~nm, with a peak at $\sim$400~nm, which is reasonably well-matched to the emission of BGO in $\sim$370-650~nm, with the peak at $\sim$480~nm. 

A 1~mm thick Eljen Technology EJ-560 silicone pad and EJ-550 optical grease couple the PMT to the scintillator and protect the thin PMT window from vibrations and shocks. 
The pads are kept in compression using a phosphor-bronze spring mounted behind the PMT.

\begin{figure*}[tb]
    \centering
    \includegraphics[width=0.9\textwidth]{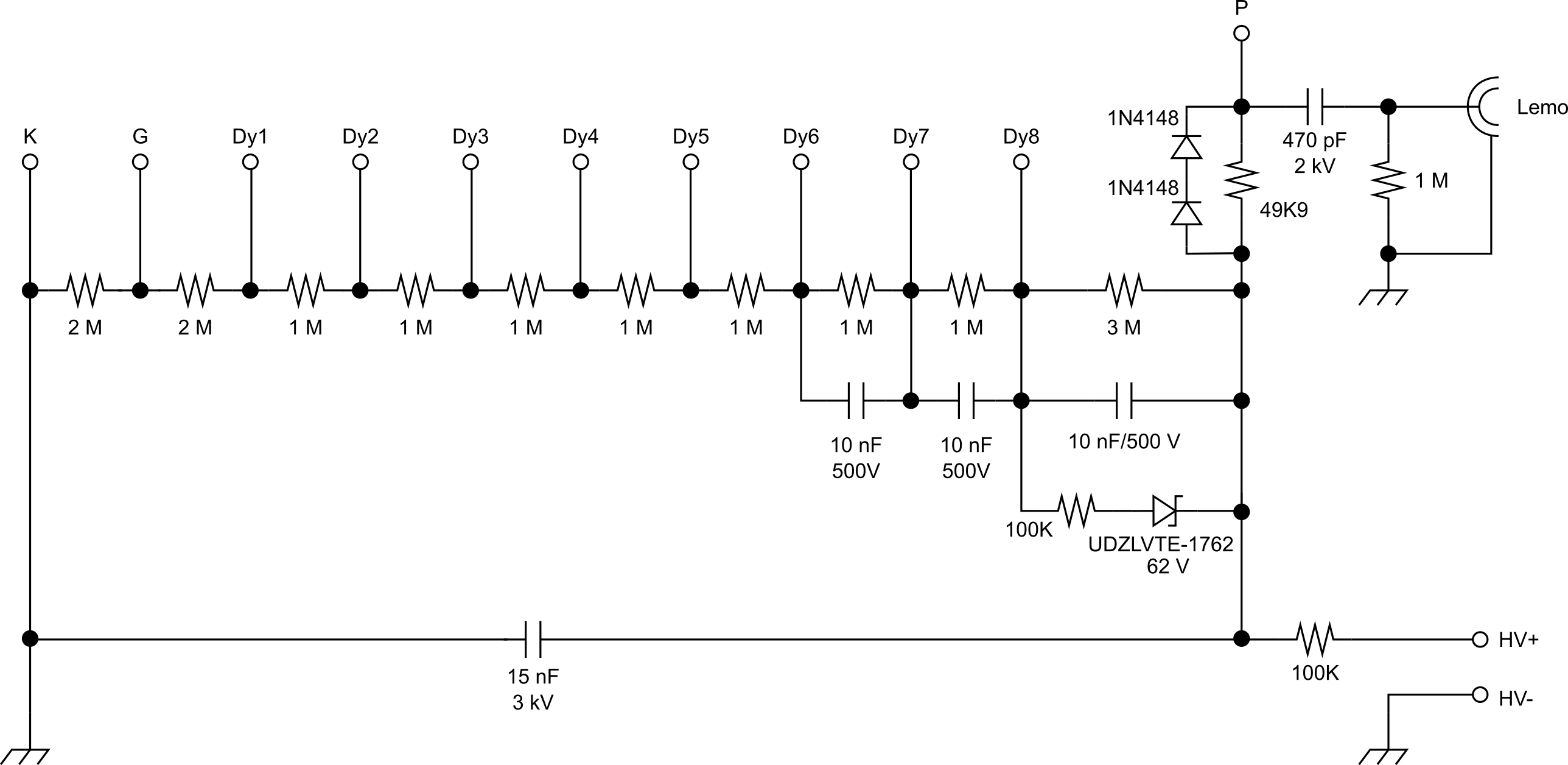}
    \caption{\label{fig:PMT-bleeder} Resistive divider chain for supplying PMT voltages. The inclusion of a 62~V Zener diode in the final dynode stage ensures that large energy deposits do not lead to saturation of subsequent readout electronics. Legend --  K: PMT cathode; Dy: dynode; G: grid; P: anode.
The total current drawn by the divider is 122~$\mu$A.}
\end{figure*}

\begin{figure}[tb]
\begin{center}
    \includegraphics[width=0.5\textwidth]{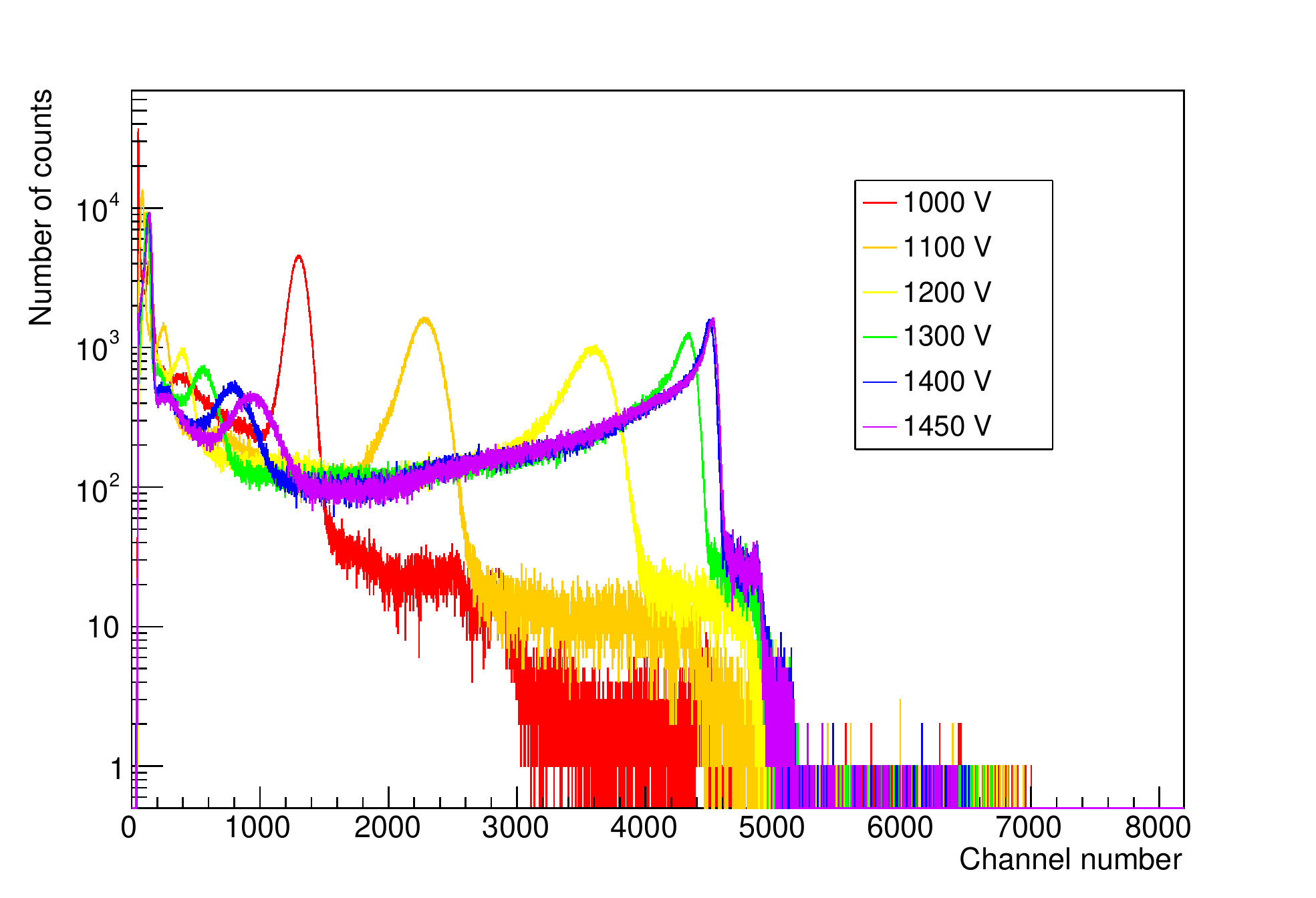}
    \includegraphics[width=0.5\textwidth]{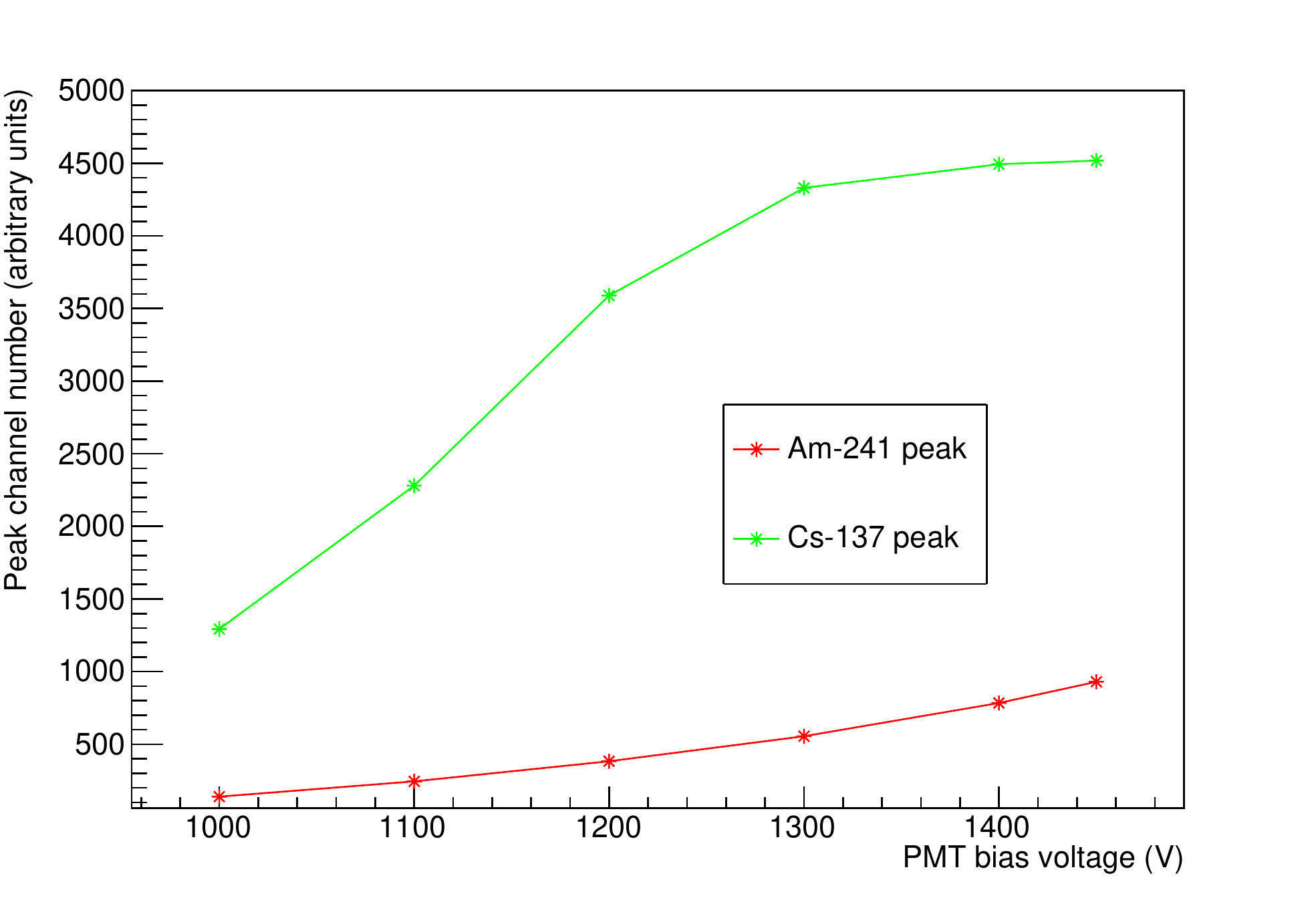}
\end{center}
\caption{\label{fig:clamping} {\bf Top:} Shield pulse height spectra for \mbox{$^{241}$Am} and \mbox{$^{137}$Cs} for different PMT bias voltages. As the voltage increases, the pulse amplitude for a given energy deposit increases, until the clamping limit is reached. 
{\bf Bottom:} Positions of the \mbox{59.5 keV} and \mbox{662 keV} peaks (from \mbox{$^{241}$Am} and \mbox{$^{137}$Cs}, respectively) as a function of PMT bias voltage. The high-energy photons from \mbox{$^{137}$Cs} are suppressed due to the clamping, while the low-energy events from \mbox{$^{241}$Am} are not affected.}
\end{figure}

The PMT was biased as shown in Figure~\ref{fig:PMT-bleeder}. At a representative operating voltage of 1400~V, the photomultiplier gain exceeds 10$^6$. 
To limit the shield dead-time resulting from large-amplitude pulses (requirement~\ref{req:deadmitigate}), a Zener-diode clamp is added to the final stage of the dynode chain, following the approach detailed in~\cite{Tanihata.1999}. 
The effect of this is demonstrated in Figure~\ref{fig:clamping}. Components are mounted on a double-sided custom printed circuit board (PCB) which is soldered onto the PMT leads. Signal connections to the PCB are made using 50~$\Omega$ LEMO 00-series connectors.
When designing the PCB, care was taken to minimise the electric field gradient between components and/or traces. 
Teledyne Reynolds 600-series connectors and factory-terminated cable assemblies are used for high-voltage connections. 
The bias voltage for each PMT is provided by a unique DC/DC unit (HVM Technology UMHV1220), which can provide a maximum 250~$\mu$A at 2000~V from a 12~V supply. 

\subsection{Discharge prevention}
A common failure mode for balloon missions utilising high-voltage components is electrical discharge in the rarefied atmosphere present at float altitude (3-5~hPa). 
As described by Paschen's Law~\cite{Keidar.2018}, the potential difference required to initiate electrical discharge between two conductors in a gas depends on the conductor separation and pressure. In air at mbar-level pressure, the discharge voltage can be as low as a few 100~V for conductors separated by a few~mm. All high-voltage components are therefore encapsulated in a two-component silicone potting compound, Momentive RTV627. The opaque compound has a dielectric strength of 510~V/mil and its low viscosity reduces the risk of forming voids inside the encapsulation. 
Prior to potting, solder joints were inspected to confirm that no sharp protrusions were present. All surfaces which come into contact with the potting compound were treated with Momentive SS4155 primer compound to improve adhesion.
The PMT/HV divider assembly is inserted into a plexiglass tube, with the window-end of the tube sealed against the PMT to allow the remaining tube volume to be filled with potting compound.
The mixed potting compound is placed in a degassing chamber at a vacuum of $\sim$1~mbar for $\sim$2~minutes to expel air bubbles. The mixture is then slowly poured into the PMT assembly with care taken to avoid the formation of voids. Once pouring is completed, the assembly is again placed into the degassing chamber under a vacuum of $\sim$1~mbar for $\sim$20~minutes. The assembly is then placed in a levelled oven and left to cure at 40$^\circ$C for 48~hours.  
The same procedure was used for potting the high voltage DC/DC converters, which are individually housed in aluminium enclosures mounted on the polarimeter electronics stack.

The potted PMTs are covered in 0.15~mm thick high-permeability magnetic shielding foil (Thorlabs MSFHP) to shield the PMTs from external magnetic fields. 
The PMT assembly is covered in opaque aluminium tape to prevent ambient light leakage into the PMT tube. 

\subsection{Data acquisition} 
\label{sec:daq}
\begin{figure*}[tb]
    \centering
    \includegraphics[width=\textwidth]{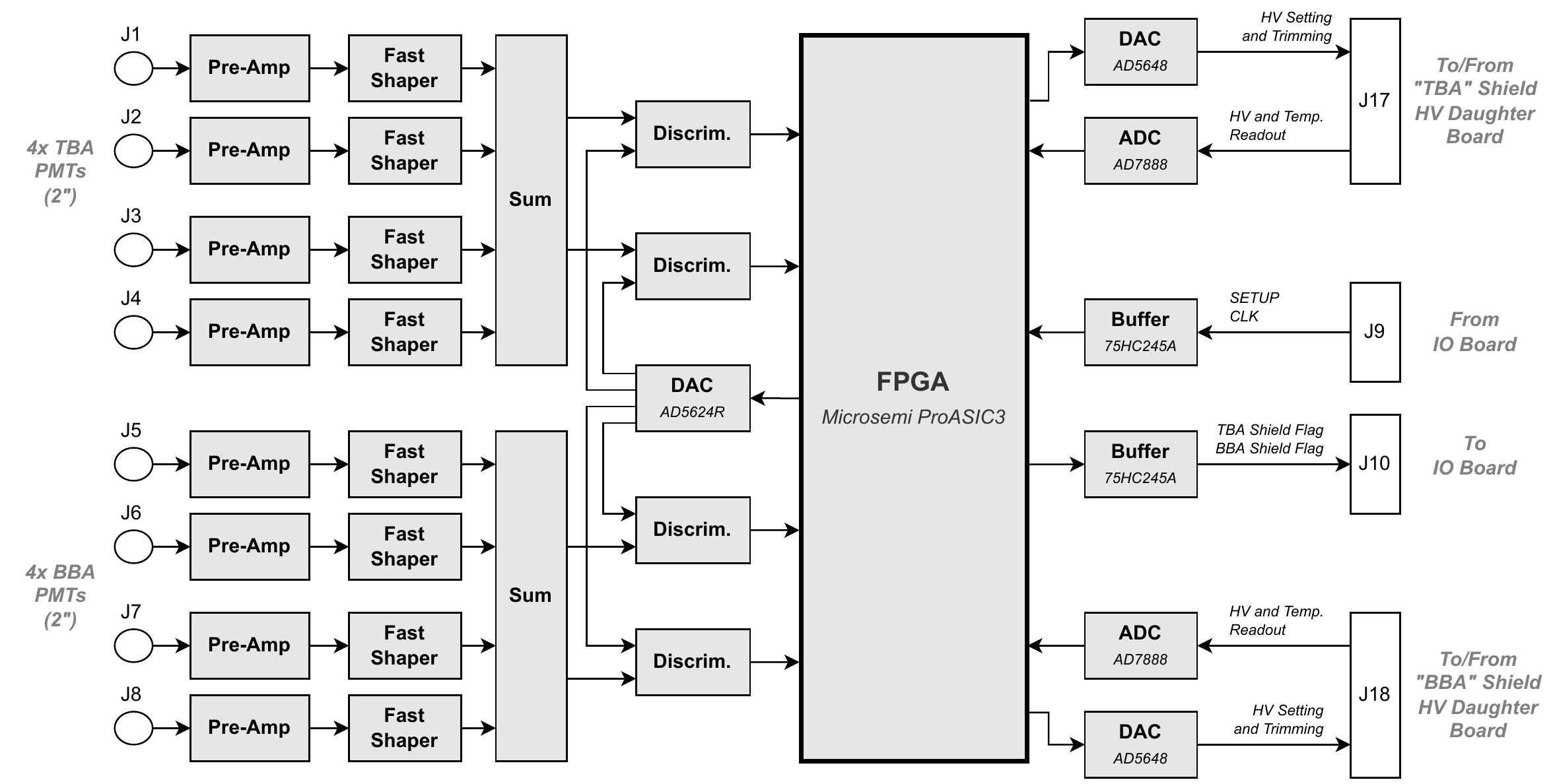}
    \caption{\label{fig:dacq}The shield data acquisition system. 
    The 8 shield PMT outputs are fed into the board and a shield veto flag for each of TBA and BBA is sent to the polarimeter readout.}
\end{figure*}

The design of the shield data acquisition (DAQ) system is based on that developed for \textit{X-Calibur}. A block diagram is shown in Figure~\ref{fig:dacq}. 

The preamplifier feedback resistor and shaper pole-zero resistor are chosen so that the decay time of the PMT signal fed to the discriminator stage is minimised, without generating a significant undershoot component. When combined with the effect of PMT pulse clamping, the shield dead-time is reduced to a few~$\mu$s (compared to 50~$\mu$s for {\it X-Calibur}), thereby allowing a maximum veto rate of a few hundred kHz (requirement~\ref{req:deadmitigate} and Section~\ref{sec:ledratetest}).
To provide more flexibility when setting up the anticoincidence veto, the FPGA logic allows the width and delay of the shield veto pulse to be adjusted, as well as the option to retrigger the veto pulse if a shield discriminator fires while the veto is asserted.
Additional FPGA modifications allow the shield performance to be studied during flight. Counters are implemented to measure the discriminator rate, veto signal rate and corresponding veto time (requirements~\ref{req:vetorate}, \ref{req:thrscan} and \ref{req:dtime}).

\begin{figure}[tb]
    \centering
    \includegraphics[width=0.5\textwidth]{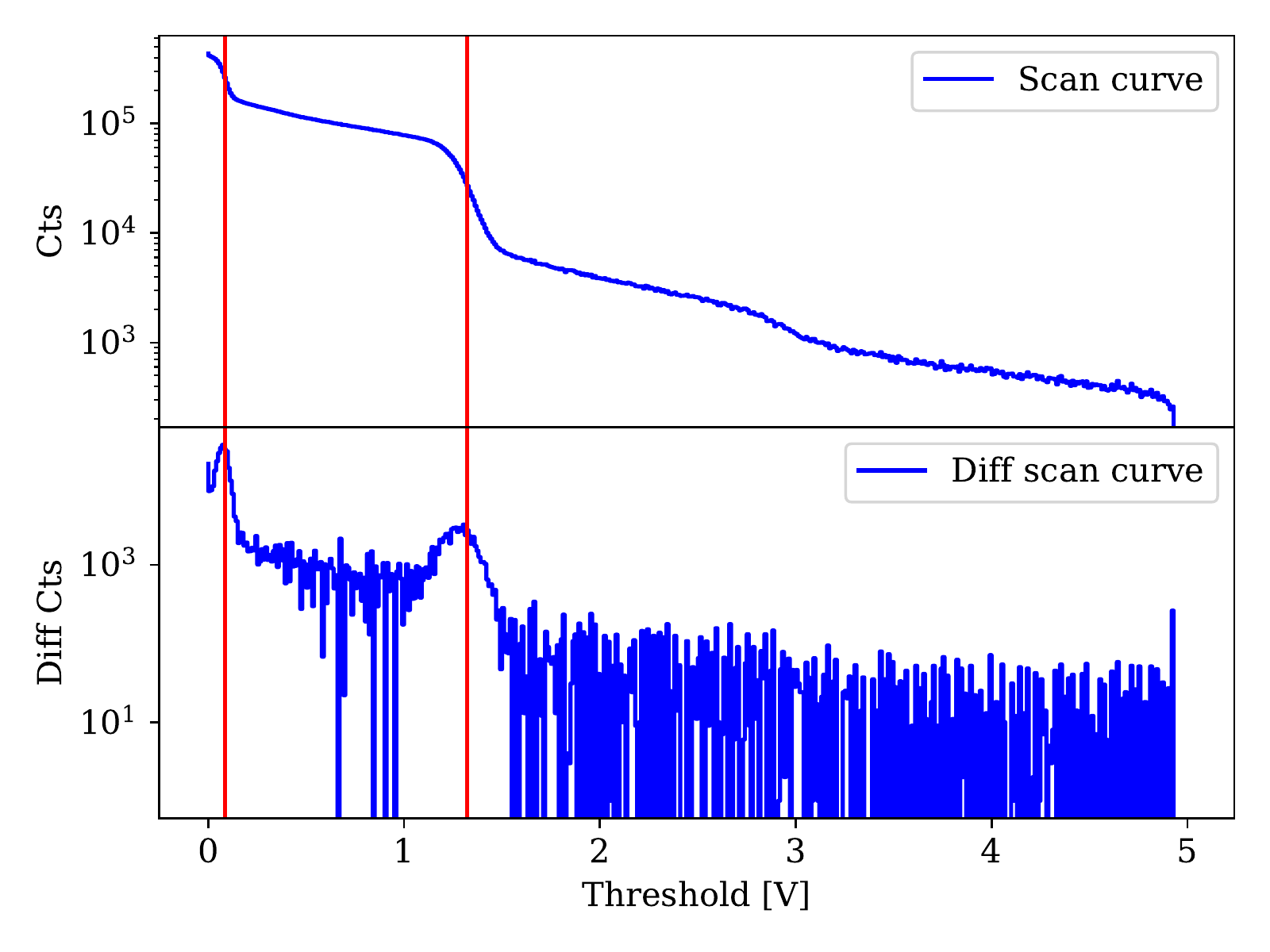}
    \caption{\label{fig:threshscan}. A 40-minute long threshold scan for TBA with a \mbox{$^{137}$Cs} and \mbox{$^{241}$Am} radioactive source line positions marked. The shoulder at $\sim$3~V threshold is due to $^{40}$K decay, as discussed in Section~\ref{sec:background}.}
\end{figure}

As shown in Figure~\ref{fig:dacq}, the summed signal from TBA and BBA is fed to two discriminators. One generates shield veto signals, while the other one is used for a newly-implemented threshold scan. 
By scanning through the range of threshold settings, a cumulative counts spectrum can be derived. 
Differentiating the cumulative number of counts bin-by-bin allows an energy spectrum to be reconstructed. 
During flight, the 8-minute duration scan covers the range 0--5~V, with a step size of 50~mV. 
Data is collected for 5~s at each step.   
An example threshold curve with the shield illuminated by radioactive sources on-ground is shown in Figure~\ref{fig:threshscan}. 
In flight, the interaction of cosmic-ray, \comment{atmospheric}, or locally-produced positrons with the payload materials produces a 511~keV annihilation line, which can be used to monitor the shield response. 

%
%
\section{Qualification testing}
\label{sec:qualtest}
%

%
\subsection{Introduction}
In order to reduce the risk of in-flight failures due to high-voltage discharge, the PMT assemblies and DC/DC units were subjected to thermal cycling tests at 4~mbar vacuum (requirement~\ref{req:thermal} and requirement~\ref{req:vac}). The electronics boards comprising the data acquisition system were tested previously for the {\it X-Calibur} mission. 
The assembled shield was qualified in a thermal cycling chamber at atmospheric pressure. As well as characterising the temperature dependence of the shield response to X-rays, the test also confirms that thermal expansion stresses do not affect the photomultiplier optical coupling or the ambient light seal. 
A long-term study of the shield response was also conducted with the shield operated in a larger vacuum chamber at room temperature. 

%
\subsection{Thermal cycling of high-voltage components in vacuum}
Devices under test were placed within a vacuum chamber housed inside a thermal cycling chamber. 
High-voltage (1450~V) was supplied to the PMTs using a CAEN N472 4-channel power supply, with connections made through the wall of the pressure vessel. 
The unit provides a current monitor output, which was used to detect discharge events. 

\begin{figure*}[tb]
\begin{center}
    \includegraphics[width=0.9\linewidth]{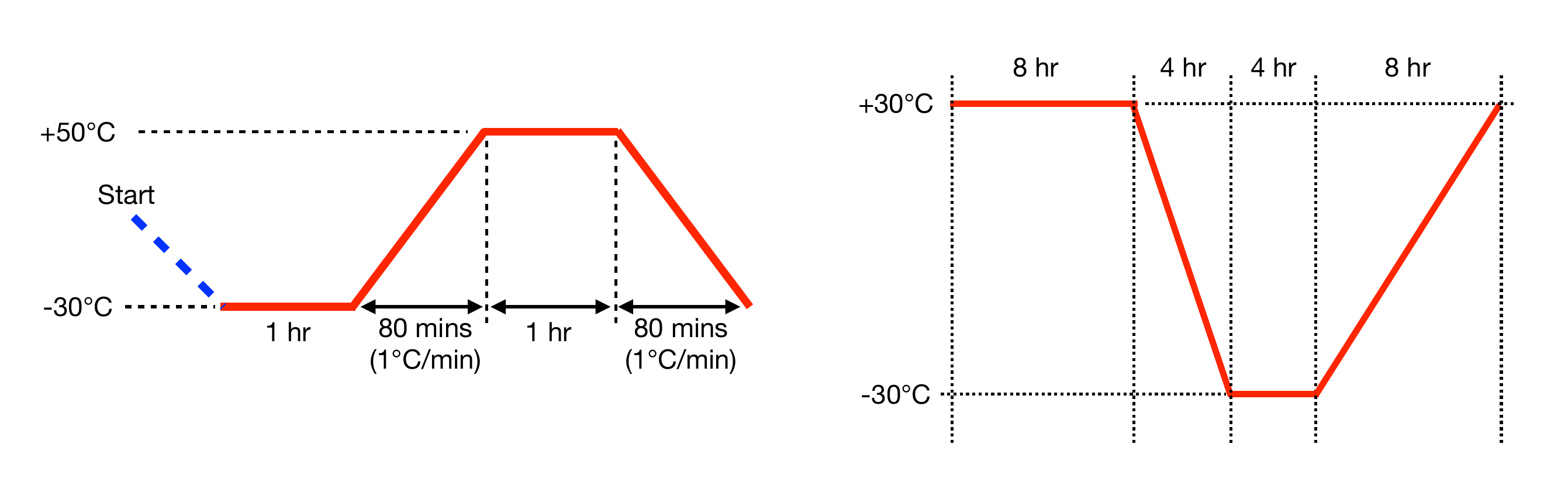}
\end{center}
\caption{\label{fig:thermal_cycle} {\bf Left:} The thermal cycling profile used during the 24-hour long 'stress-test' of components in order to reveal possible workmanship issues. {\bf Right}: The thermal cycling profile used during the week-long "flight-test" of components. 
}
\end{figure*}

The potted PMT assemblies were subjected to two types of thermal cycling test at 4~mbar vacuum. 
The first test was designed to address possible workmanship issues during fabrication of the PMT assemblies. This 'stress-test' comprised 5 cycles and lasted for 24~hours, with the temperature profile shown in Figure~\ref{fig:thermal_cycle}. The fast ramping times are not representative of flight conditions, but were chosen to provoke component failure due to, e.g., differential thermal expansion effects. A single-photoelectron reference spectrum was recorded before and after the test for each PMT. No significant differences were seen between these spectra and no discharge events were registered for any of the PMTs during the cycling.
Each PMT assembly was visually inspected at the end of the test. Although the potting compound was sometimes found to partially detach from the plexiglass tube after thermal cycling, the bulk of the potted volume was not affected.

The second thermal cycling test was designed to replicate conditions expected during a balloon flight from Esrange to Canada. 
Temperature measurements taken during a comparable flight of the {\it PoGO+} mission in summer 2016~\cite{Chauvin.2017} were used to define the temperature cycle profile (Figure~\ref{fig:thermal_cycle}). Temperatures were measured inside the {\it PoGO+} gondola, so that direct solar illumination is avoided -- as is the case for the {\it XL-Calibur} shield. 
This test was conducted over 7 cycles of 24 hours, corresponding to the expected flight time. No discharge events were registered during the test. The single-photoelectron spectra measured before and after the test also showed no significant differences.

The potted DC/DC units were tested in a similar manner. In this case, low-voltage control signals were passed into the vacuum chamber, and the high-voltage output was monitored for discharge. 
After the thermal-cycling "stress-test", two of the twelve DC/DC units tested exhibited hairline cracks in the exposed surface of the potting compound. No discharges were registered for these units, but they were rejected for flight as a precautionary measure. Another unit triggered the discharge monitor continuously during the following thermal-cycling "flight-test" and was similarly rejected. The reason for the failure of this unit could not be determined.
All other units passed both thermal-cycling tests. For each functioning unit, the input/output linearity was measured before and after the test. No significant differences were observed.
%
\subsection{Thermal testing of the assembled shield}
\begin{figure}[tb]
\centering
\includegraphics[width=0.5\textwidth]{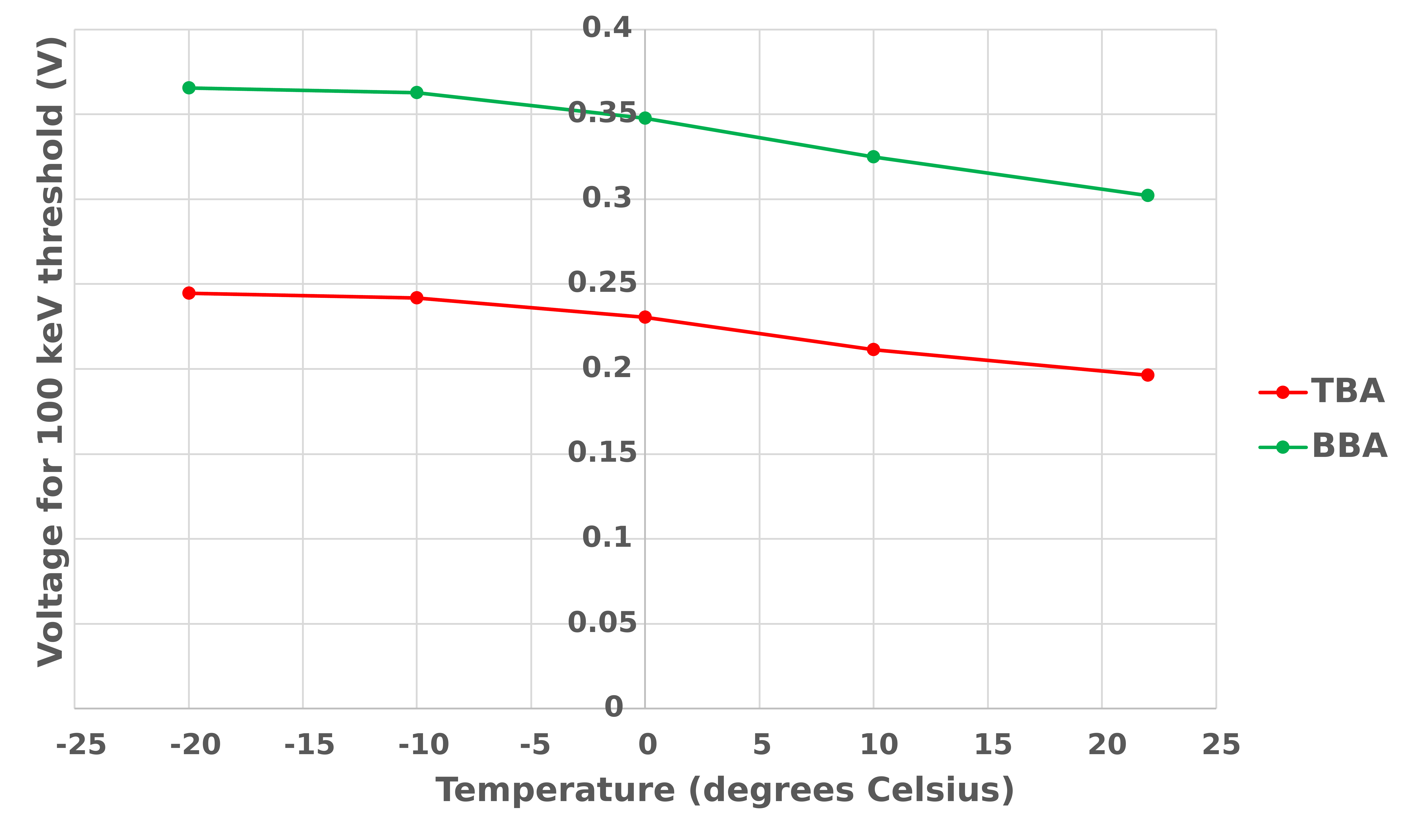}
\caption{\label{Threshold for 100 keV}Discriminator reference voltage required for setting a \mbox{100 keV} veto threshold as a function of temperature in the TBA (red) and BBA (green). The higher threshold values of the BBA result from the better light-collection efficiency of the smaller scintillator volume.}
\end{figure}

The assembled shield and electronics board were placed in the thermal chamber and threshold scans conducted at 
$-$20~$^\circ$C, $-$10~$^\circ$C, 0~$^\circ$C, +10~$^\circ$C and +25~$^\circ$C with the shield exposed to a \mbox{$^{137}$Cs} radioactive source.
The light-yield of BGO scintillator is inversely proportional to temperature ($\sim$-1\%/$^\circ$C~\cite{Wang.2014a2}).
The position of the 662~keV photopeak was used to calibrate the shield thresholds for a given temperature.
Threshold scans were acquired alternately with the shield being operated in darkness and during illumination by strong lighting at each test temperature. No significant differences were observed between these scans, showing that the mechanical structure remained light-tight.
The temperature dependence of the 100~keV threshold point
is shown in shown in Figure~\ref{Threshold for 100 keV}.

%
\subsection{Operation of the assembled shield in vacuum}
The assembled shield and electronics boards were placed inside a vacuum tank, which was pumped down to 4~mbar. 
The performance of the shield was periodically monitored during 1~week using threshold scans with the shield was exposed to $^{241}$Am and $^{137}$Cs radioactive sources. The position of both photopeaks did not change during the test, showing that the shield response was not affected by vacuum conditions.

%
\subsection{Energy threshold calibration and response uniformity}
\label{sec:threshold}
The discriminator threshold is related to deposited energy by irradiating the shield with  
\mbox{$^{137}$Cs} and \mbox{$^{241}$Am} sources, while the summing output (Figure~\ref{fig:dacq}) is connected to an external preamplifier-shaper-multichannel analyzer set-up.
The sources were placed on the symmetry axis of each shield assembly, equidistant from each PMT. 
Since the PMT gain and PMT-scintillator coupling varies, bias voltages were chosen per PMT to produce the same photopeak position. 
As a result, the shield response becomes independent of source position.

%
\subsection{Shield trigger rate}
\label{sec:ledratetest}
Two fibre-coupled 100~mW LEDs (470~nm) are connected through a fibre-splitter, with
one LED pulsed to emulate a MIP-like energy deposit, while the other mimics a \mbox{100 keV} energy deposit (requirement~\ref{req:thresh}). The smallest difference in arrival times
at which a \mbox{100 keV} pulse generates a veto, while riding on the offset baseline of a preceding large-amplitude MIP pulse, was determined to be \mbox{$\Delta\!t \lesssim$2~$\mu$s}, 
\comment{corresponding to background rate of 500 kHz impinging on the shield.}
This is much larger than the rate of background events expected in flight (requirement~\ref{req:vetorate}). 

%
\section{Background rejection performance in flight}
\label{sec:flightResults}

{\it XL-Calibur} was launched from the Esrange Space Centre on July 11$^{\mathrm{th}}$ 2022 at 23:45~UTC. The float altitude of $\sim$39.6~km was reached at $\sim$03:45~UTC the following day. 
The payload was cut from the balloon on July 18$^{\mathrm{th}}$, and landed by parachute in the Canadian Northern Territories, $\sim$500~km Northwest of Yellowknife at ~07:40 UTC.
The payload suffered little visible damage due to the marshy conditions at the landing site. Hardware is being returned to the laboratory for testing and refurbishment before the next balloon flight. 

The shield operated as expected during the flight, with the exception of the following two events: $(1)$ approximately 1.5~hours after launch, at an altitude of $\sim$20~km, one of the BBA PMTs was turned off due to an anomalously low high-voltage reading (indicative of discharge); $(2)$ during the early hours of July 15$^{\mathrm{th}}$, one of the TBA PMTs became noisy and was also switched off. The reason for these failures will be determined when hardware is inspected. The effect on shield performance is discussed below. 

\begin{figure*}[h]
    \centering
    \includegraphics[width=\textwidth]{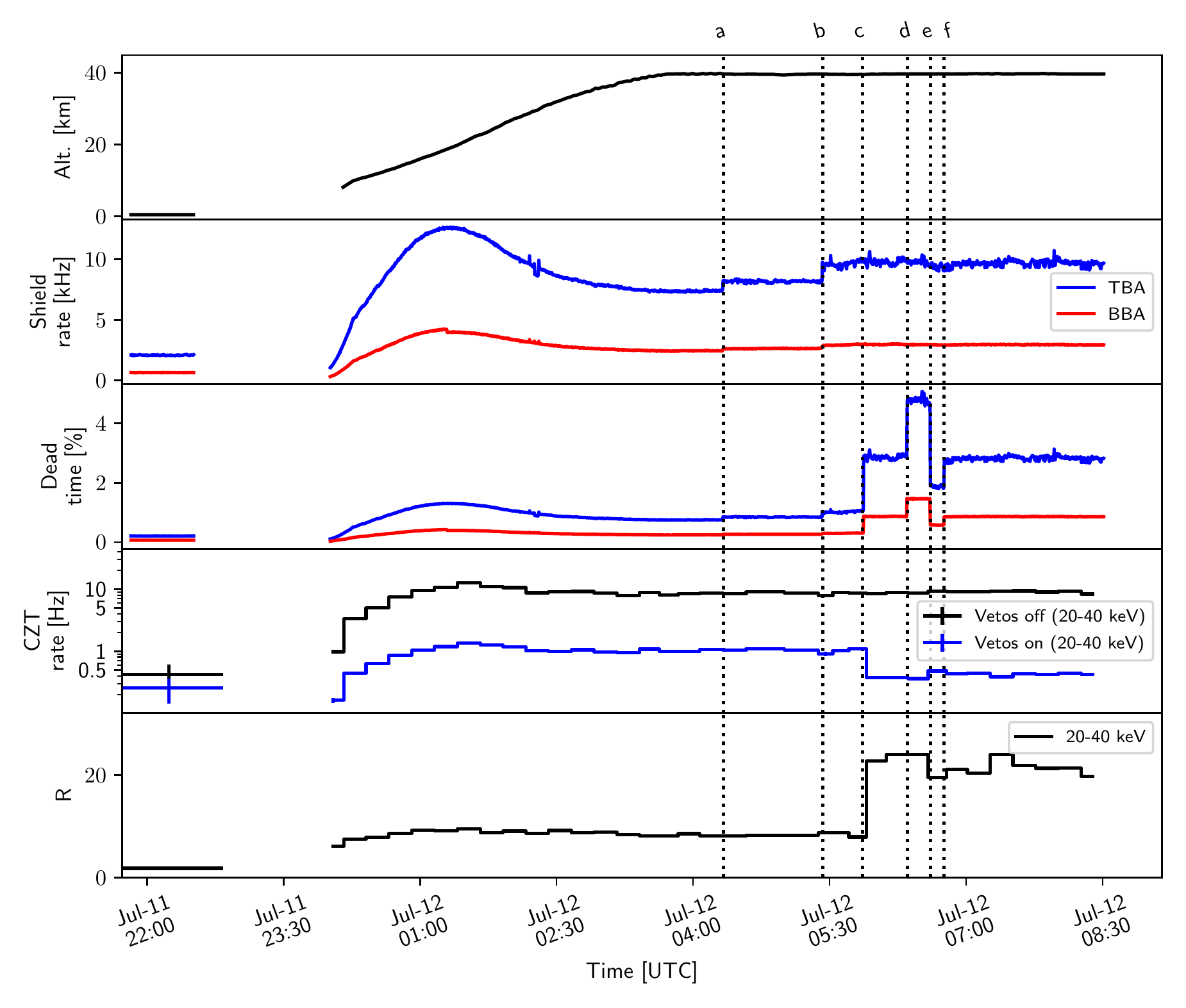}
    \caption{From top to bottom: balloon altitude, shield discriminator rate, shield dead-time fraction, polarimeter (all CZT) 1-hit rate, veto efficiency ratio (R=vetos-off/vetos-on polarimeter 1-hit rate). The leftmost data-set was recorded before launch, while the rightmost data-set is post-launch. In the discriminator rate panel, the small step in the BBA rate (red curve) in the vicinity of the Regener-Pfotzer maximum corresponds to the failure of a BBA PMT. During ascent, discriminator levels were set to achieve a 100~keV threshold, and the veto window with was 1~$\mu$s, as motivated by ground-tests. Once at float altitude, the shield configuration was adjusted as follows (threshold, width): {\it (a)} (75~keV, 1~$\mu$s); {\it (b)} (50~keV, 1~$\mu$s); {\it (c)} (50~keV, 3~$\mu$s); {\it (d)} (50~keV, 5~$\mu$s); {\it (e)} (50~keV, 2~$\mu$s); {\it (f)} (50~keV, 3~$\mu$s). 
    \label{fig:rates}}
\end{figure*}

\begin{figure}[h]
    \centering
    \includegraphics[width=0.5\textwidth]{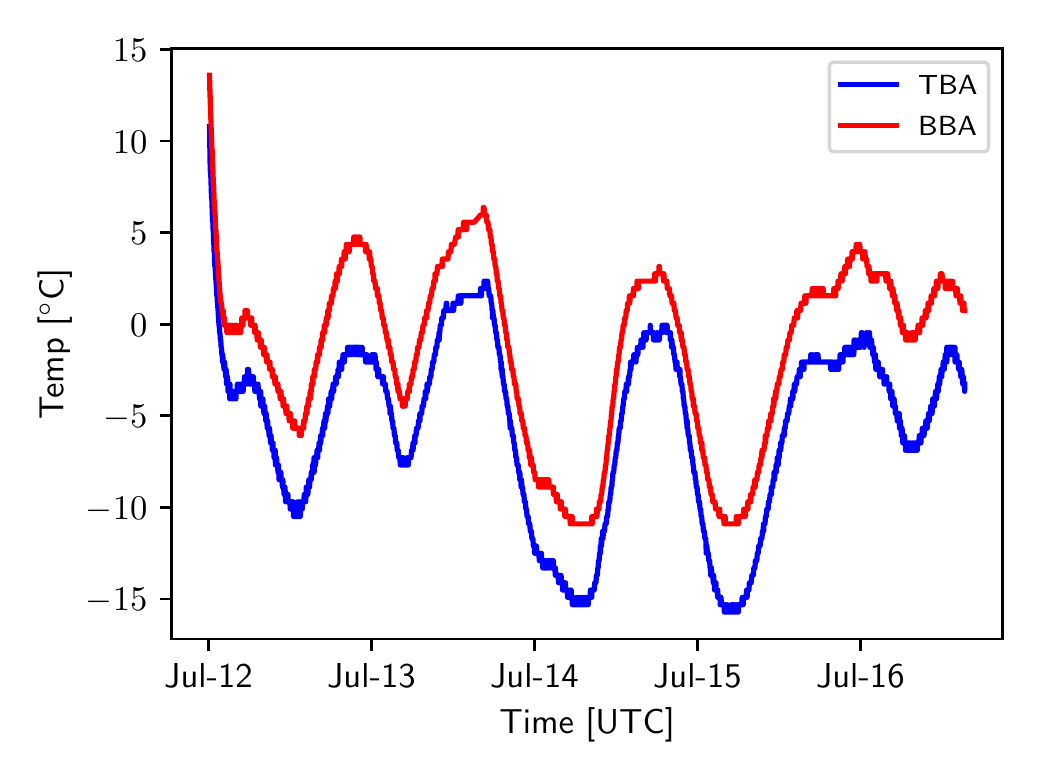}
    \caption{The temperature of the BBA and TBA mechanical structures during the flight. 
    After ascent to float on July 12th, the large variations are due to day- and night-time conditions. Day-to-day differences arise due to altitude variations, the pointing direction, and whether the balloon was flying over sea or land, e.g. the colder night-time temperatures on July 14th and 15th correspond to transits over Iceland and Greenland, respectively.   
    The polarimeter is in thermal contact with BBA, which consequently is systematically warmer.
    \label{fig:ShieldTemperature}}
\end{figure}

The shield discriminator rate and dead-time fraction are shown in Figure~\ref{fig:rates} for the ascent phase and during commissioning at float altitude. The maximum rate of particles detected by the shield ($\sim$16~kHz) occurs at the Regener-Pfotzer maximum, for a corresponding \comment{shield} dead-time fraction of $<$2\%. Once float altitude was reached, the shield veto threshold and veto width were adjusted to optimise the shield background rejection performance. A shield threshold of $\sim$50~keV was set, without introducing significant dead-time. Increasing the veto width to 3~$\mu$s (1~$\mu$s was used during on-ground testing) was found to significantly reduce the 1-hit CZT rate.   
This is due to the longer BGO scintillation decay time~\cite{Gironnet.2008} resulting from the lower average shield temperature (see Figure~\ref{fig:ShieldTemperature}) compared to on-ground conditions. The lower temperature also increases the BGO light-yield.   

\begin{figure}[h]
    \centering
    \includegraphics[width=0.5\textwidth]{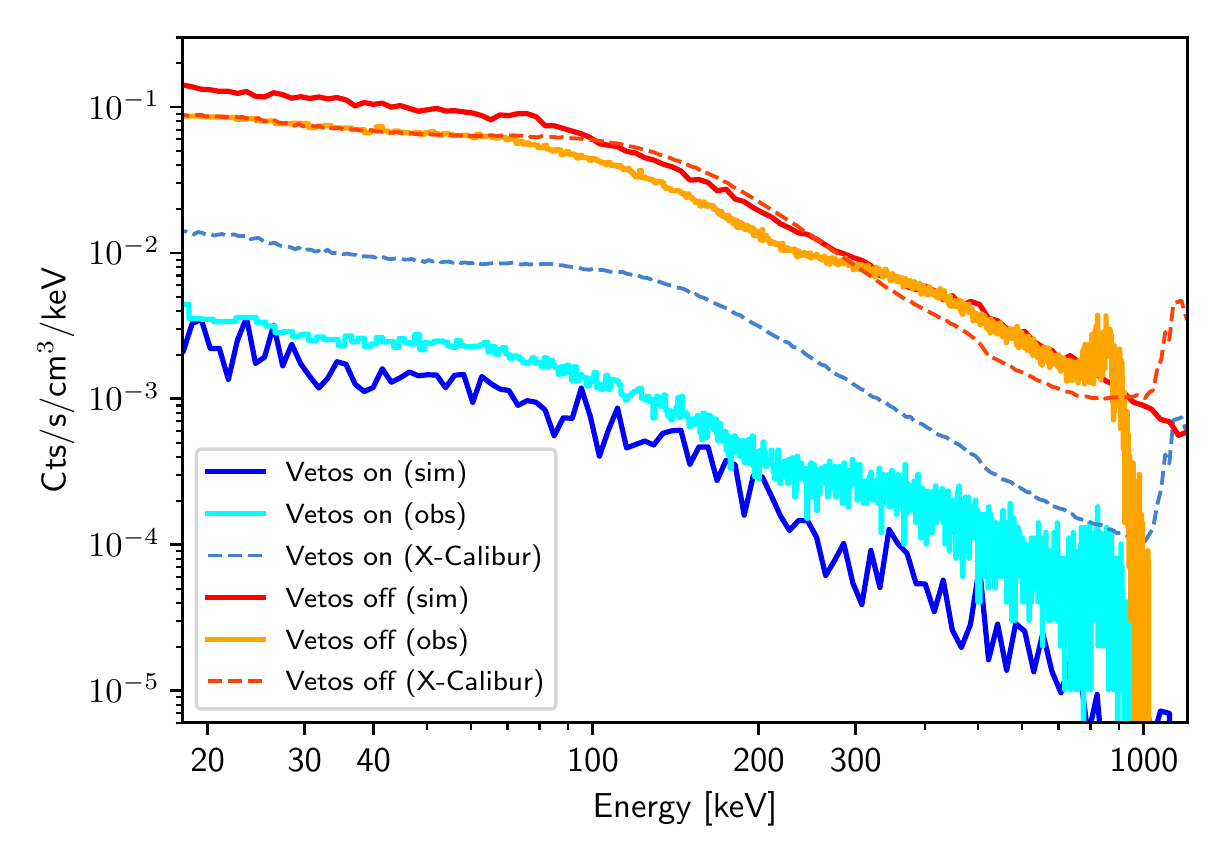}
    \caption{A comparison between the simulated and observed 1-hit polarimeter (all CZT) energy spectrum for shield vetos-on (lower two solid curves) and shield vetos-off (upper two solid curves). The peak in the observed spectra at $\sim$1~MeV is an electronics artefact. Energy deposits in TBA and BBA have been scaled by 75\% to account for the malfunctioning PMTs. The dotted curves show the corresponding background spectra (vetos-on and vetos-off) measured by {\it X-Calibur} during the 2018 Antarctica flight. Note that $y$-axis values are quoted as cm$^{-3}$ to allow the missions to be compared.
The improved background rejection performance of {\it XL-Calibur} is clearly evident.  
    \label{fig:CZTRates}}
\end{figure}

Since each shield scintillator is read out with multiple PMTs, the failure of a single read channel had limited effect. 
The effective shield energy threshold increases, e.g. the $\sim$50~keV threshold set during flight was closer to $\sim$70~keV. 
It is also possible that a position-dependence is introduced to the shield response, but subtracting background measured in off-source pointings would mitigate this.   
%
These final settings yielded a shield rate of $\sim$12~kHz, for a \comment{shield} dead-time fraction of $\sim$4\%.      
The rate of 1-hit events measured for all CZT detectors (20--40~keV)\footnote{This is the energy range with best signal-to-background performance for flight observation conditions.} is 8.2~Hz for vetos-off, and 0.5~Hz for vetos-on. 
The CZT 1-hit rates predicted by the simulation are 10.9~Hz (vetos-off) and 0.16~Hz (vetos-on). 
In all cases, the error on the measured rate is negligible.
Corresponding energy spectra are shown in Figure~\ref{fig:CZTRates}. 
There is reasonable agreement between data and simulation with vetos-off, indicating that the flux of background particles incident on the shield is well-modelled. 
\comment{The background-rejection efficiency (vetos-on) is over-estimated in the simulation. 
This is ascribed to simplifications in the Geant4 model, as described in Section~\ref{sec:simulation}.}  

\begin{figure}[h]
    \centering
    \includegraphics[width=0.5\textwidth]{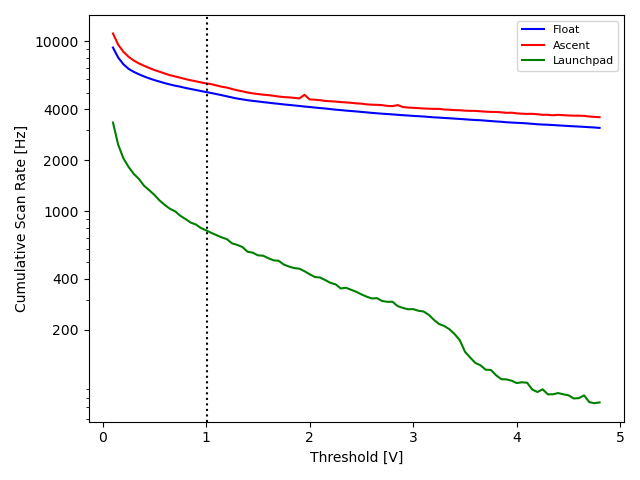}
    \caption{Thresholds scans (TBA) acquired on-ground, during ascent and at float altitude. The feature seen at a threshold voltage of $\sim$3.5~V is due to $^{40}$K decays ($\sim$1.5~MeV). 
    The differential scan in the vicinity of the 1 V range (dotted line) is shown in Figure~\ref{fig:thscan_511}.
        \label{fig:thscan_comp}}
\end{figure}

\begin{figure}[h]
    \centering
    \includegraphics[width=0.5\textwidth]{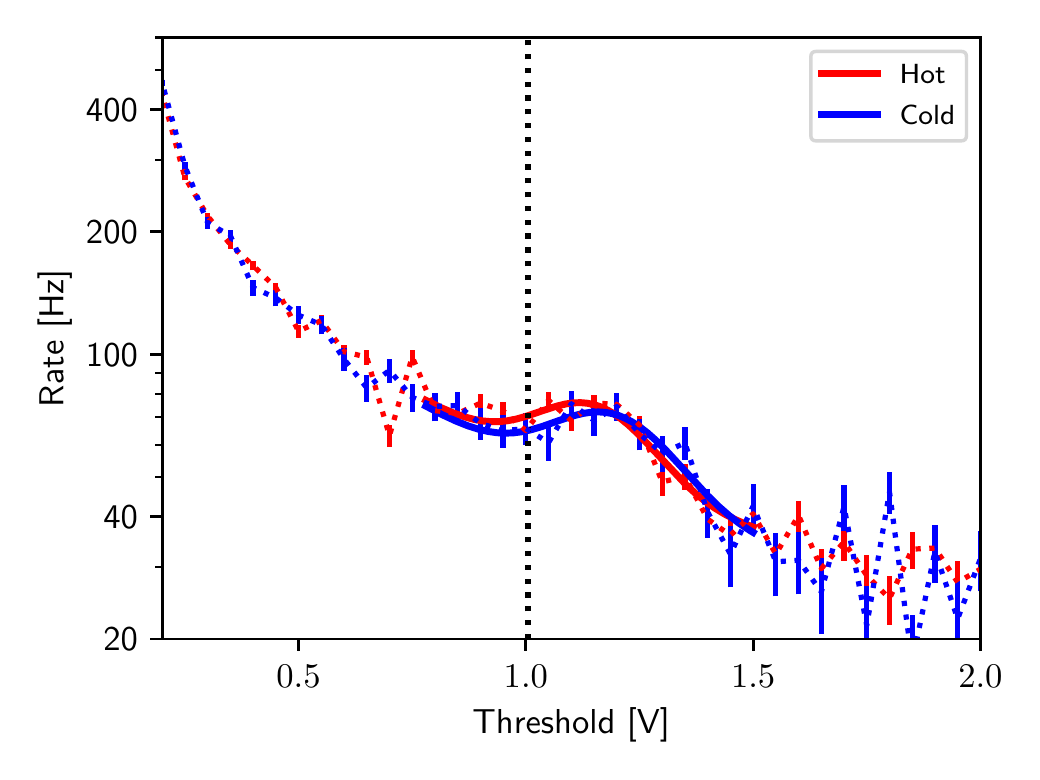}
    \caption{The differential threshold scan around the location of the 511~keV position annihilation peak. The vertical dotted line shows the expected peak position from ground-tests conducted at room temperature. The red (blue) curve shows threshold scans taken during day-time (hot) and night-time (cold) conditions, as shown in Figure~\ref{fig:ShieldTemperature}.       
    \label{fig:thscan_511}}
\end{figure}

Threshold scans taken on ground, during ascent and at float altitude are shown in Figures~\ref{fig:thscan_comp} and \ref{fig:thscan_511}. The background on-ground is dominated by relatively low energy interactions compared to float altitude. The active background rejection performance at float altitude is therefore superior to that measured on-ground, as witnessed by the lower panel in Figure~\ref{fig:rates}.
The ground spectrum shows a feature at an energy of $\sim$1.46~MeV, which is due to gamma-ray decay products of the $^{40}$K, as discussed in Section~\ref{sec:background}.   
As shown in Figure~\ref{fig:ShieldTemperature}, the shield temperature followed a diurnal pattern during the flight, governed by solar heating\footnote{The temperature cycles used during qualification testing (Figure~\ref{fig:thermal_cycle}) were not well-matched to the measured shield temperature. These flight measurements will inform future qualification tests.}. 
\comment{Figure~\ref{fig:thscan_511}} shows threshold scans accumulated during day-time (average temperature $\sim$0$^\circ$C) and night-time (average temperature $\sim$$-$(5-10)$^\circ$C). 
The BGO light-yield is consequently higher than during on-ground testing
(Figure~\ref{Threshold for 100 keV}), and the 511~keV peak position moves to a higher apparent energy. 
The observation of the 511~keV peak demonstrates the potential of monitoring the shield energy scale in-flight, and the effect of variations in the shield temperature.

\begin{figure}[h]
    \centering
    \includegraphics[width=0.5\textwidth]{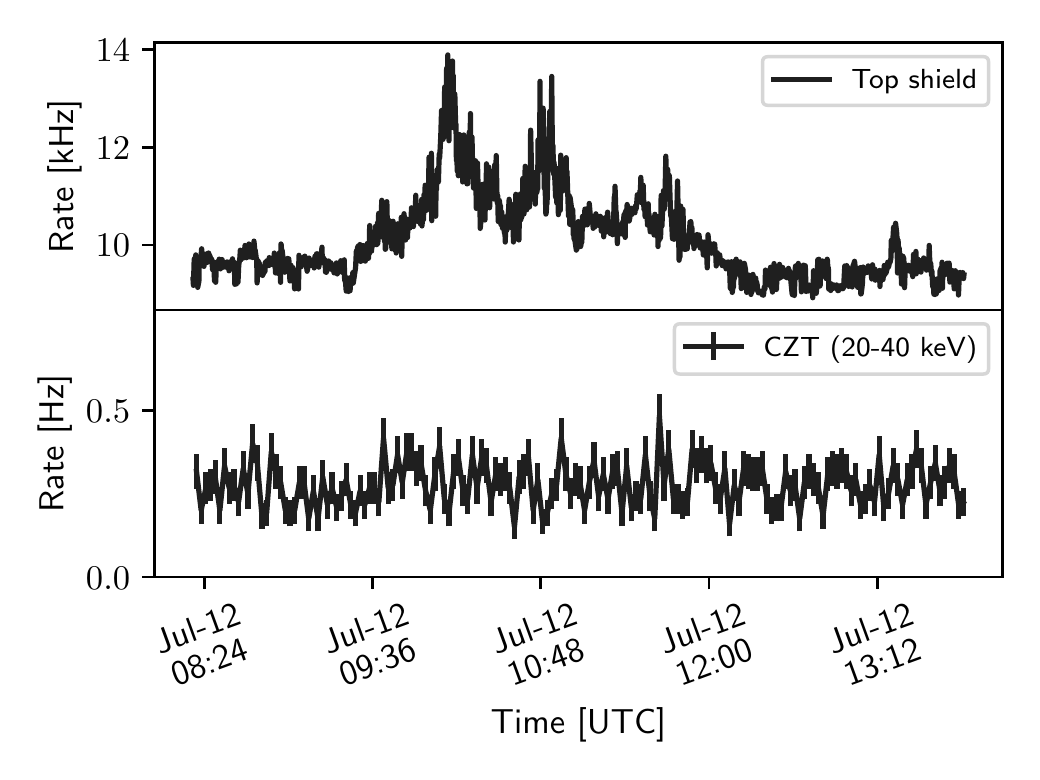}
    \caption{The discriminator rate in the anticoincidence shield (top) and CZT detector (bottom) as a function of time. In the selected period, there is a transient increase in the background rate. The polarimeter 1-hit rate remains unchanged. 
    \label{fig:trans}}
\end{figure}

Several times during the flight, the rate of particles incident on the shield was seen to significantly increase temporarily. An example is shown in Figure~\ref{fig:trans}. Such transients were seen on each day of the flight and varied in duration from hours (as shown here) to minutes. The origin of these transient events is interesting, but has not been studied in detail since the effect on the polarimeter 1-hit rate with vetos-on is negligible. Similar transient activity was previously seen in the BGO anticoincidence shield of the {\it PoGOLite} and {\it PoGO+} balloon flights from Esrange\footnote{V.~Mikhalev, Ph.D. Thesis, KTH Royal Institute of Technology (2018). 
\url{https://kth.diva-portal.org/smash/record.jsf?pid=diva2\%3A1208715}.}.

%
\section{Discussion and outlook}\label{sec:discussion}

X-ray polarimetry of celestial compact objects in the hard X-ray band is being developed through ballon-borne observations from the stratosphere. The approach is attractive since payloads can be recovered after the flight and modifications can be made based on flight experience. The background environment at the float altitude provided by current balloon technology ($\sim$40~km) presents a significant challenge when designing instrumentation. The sensitivity of measurements is largely dictated by the efficiency of the background suppression scheme adopted (Equation~\ref{eqn:MDP}).

The aim of the {\it XL-Calibur} mission is to achieve 1\%-level MDP, which requires a two orders of magnitude suppression of the polarimeter background rate to $<$1~Hz. The relatively thick BGO anticoincidence shield newly developed for {\it XL-Calibur} has been shown to meet this goal, yielding a 1-hit polarimeter background rate of $\sim$0.5~Hz (20--40~keV), measured with shield vetos applied. In Table~\ref{table:rates}, the cumulative effect of data selections on the polarimeter rate is shown. 

\begin{table}[h]
\begin{center}
\begin{tabular}{|l|c|c|}
\hline
				& \multicolumn{2}{c|}{Polarimeter rate (Hz)} \\ 
 				& Vetos-off 	& Vetos- on 	\\ 
\hline
All events 			&	150	&	35	\\
\hline
20-40 keV events		&	70	&	5	\\
\hline
1-hit 20-40 keV events	&	10	&	0.5	\\
\hline
\end {tabular}
\caption{The effect of data selections on the polarimeter background rate, where all CZT detectors are considered.}
\label{table:rates}
\end{center}
\end{table}

The simulation environment used to optimise the shield design uses Geant4 to model the shield geometry and the interactions of particles with spectra defined in the MAIRE radiation environment model (formally known as QARM). This approach has now been validated for both the {\it X-Calibur}~\cite{Abarr.2022} flight from McMurdo, Antarctica, and the {\it XL-Calibur} flight from Esrange. There is good qualitative agreement between measurements and the simulation. 
\comment{Determining the origin of the factor of $\sim$2 difference lies outside the scope of this paper, but may be the subject of future work}.

The use of high-voltage PMTs allows the BGO scintillation light to be read out with relatively high-efficiency, but introduces the risk of component failure due to high-voltage breakdown processes which may occur in the rarefied atmosphere.
Despite lengthy pre-flight qualification tests, where the environment at float was reproduced, components failed during the flight. The redundant read out scheme meant that there was negligible effect on the shield performance, but the situation would have been more precarious if multiple read-out channels had failed for a given scintillator assembly.
Post-flight testing will reveal the reason for the failures. 

\comment{For future flights, scintillator read-out based on robust solid-state photodetectors may be studied. 
The {\it Hitomi} BGO anticoincidence shield was read out using avalanche photodiodes (APD)~\cite{Nakazawa.2018}. 
The silicon photomultiplier (an array of APDs operated in Geiger mode) is a more recent development with a number of benefits over traditional PMTs, e.g.
a low operating voltage ($\sim$50-100~V) eliminates discharge-related failures, and the device are insensitive to magnetic fields.
The quantum efficiency and spectral response are similar to a PMT, and silicon photomultiplier arrays offer comparable light-collection area. 
The relatively strong temperature dependence of the gain may require more detailed on-ground characterisation and complicate flight operations. 
An increasing number of missions have adopted these devices, providing valuable experience in their use in the (near-)space environment~\cite{Mitchell.2019}, \cite{Sharma.20206d}, \cite{Kushwah.2021}, \cite{An.2022}.}

Results on source observations are pending the completion of data analysis. 
Technical issues encountered during the flight may significantly degrade the signal detection efficiency.
The full potential of the {\it XL-Calibur} design will then be revealed during 
future flights, from from McMurdo and/or Esrange~\cite{Abarr.2021}.

%
%
\section*{Acknowledgements}
{\it XL-Calibur} is funded by the NASA APRA program under contract number 80NSSC18K0264.
KTH authors acknowledge support from the Swedish National Space Agency (grant numbers 199/18 and 2020-00201), and the Swedish Research Council (grant number 2016-04929), which allowed the development and production of the anticoincidence shield.  
We are indebted to Rolf Helg (KTH) and Bosse Barks\"{a}ter (KTH) for manufacturing the mechanical parts of the shield during the challenging conditions presented by the pandemic. We also thank Polymerteknik AB for expertly gluing the BGO crystal segments. 
The majority of the thermal and vacuum test infrastructure used in this work was provided by the KTH Space Centre. 
Important technical contributions to the project have been made by 
Dana Braun, Paul Dowkontt, Garry Simburger (Washington University in St. Louis) and Victor Guarino (Guarino Engineering, Illinois).
The NASA-CSBF and Esrange teams are thanked for providing balloon launch and operation services. We also thank colleagues from NASA Wallops Flight Facility for operating the WASP pointing system and their support during the balloon campaign. 

%
\bibliographystyle{ieeetr}
\bibliography{XL-Calibur_shield}

%
%
\end{document}